\documentclass[11pt]{article}
\usepackage{jmlr2e}
\usepackage{abstract}
\usepackage{array}
\usepackage{bbding}
\usepackage{multirow}
\usepackage{adjustbox}
\usepackage{bbm}
\usepackage{float}
\usepackage{amsfonts}
\usepackage{graphicx}
\graphicspath{{graphics/}}
\usepackage{subfigure}
\usepackage{float}

\usepackage{tabularx,booktabs}
\usepackage{multirow}
\newcolumntype{Y}{>{\centering\arraybackslash}X}

\usepackage[ruled,vlined,algo2e]{algorithm2e}
\usepackage{algorithm}
\usepackage{algorithmic}
\usepackage{natbib}

\usepackage{lscape}
\usepackage{comment}
\usepackage{bm}
\usepackage[title]{appendix}
\usepackage{longtable}
\usepackage{mathtools}
\usepackage{dsfont}
\usepackage{fancyhdr}

\usepackage{stackengine}
\usepackage{textcomp}
\usepackage{stfloats}
\usepackage{verbatim}
\usepackage{xcolor}
\definecolor{darkblue}{rgb}{0.0,0.5,0.5}
\definecolor{blue}{rgb}{0.0,0.0,1}

\usepackage[colorlinks]{hyperref}
\hypersetup{colorlinks,breaklinks,linkcolor=blue,urlcolor=blue,anchorcolor=blue,citecolor=blue}
\usepackage{cleveref}
\usepackage{booktabs,caption}
\usepackage{multirow}
\usepackage{lscape}
\usepackage{epstopdf}
\usepackage{lineno}
\usepackage{microtype}

\usepackage{graphicx}
\usepackage{subfigure}
\usepackage{booktabs} 
\usepackage{bbm}

\usepackage{amsfonts}
\usepackage{xurl}
\usepackage{stackengine}
\usepackage{tikz}
\usetikzlibrary{decorations.pathreplacing}
\usetikzlibrary{positioning,arrows.meta,quotes}
\usetikzlibrary{shapes,snakes}
\usetikzlibrary{bayesnet}
\tikzset{>=latex}
\tikzstyle{plate caption} = [caption, node distance=0, inner sep=0pt, below left=5pt and 0pt of #1.south]
\usepackage[normalem]{ulem}
\usepackage{multirow}
\usepackage{makecell}
\newcolumntype{Y}{>{\centering\arraybackslash}X}

\firstpageno{1}

\begin{document}

\title{Bayesian inference of time-varying origin-destination matrices from boarding and alighting counts for transit services}

\author{\name Xiaoxu Chen \email xiaoxu.chen@mail.mcgill.ca\\
       \addr Department of Civil Engineering\\
       McGill University\\
       Montreal, QC  H3A 0C3, Canada
       \AND
       \name Zhanhong Cheng \email zhanhong.cheng@mcgill.ca \\
       \addr Department of Civil Engineering\\
       McGill University\\
       Montreal, QC  H3A 0C3, Canada
       \AND
       \name Lijun Sun\thanks{Corresponding author.} \email lijun.sun@mcgill.ca \\
       \addr Department of Civil Engineering\\
       McGill University\\
       Montreal, QC  H3A 0C3, Canada}

\editor{}

\maketitle

\begin{abstract}
Origin-destination (OD) demand matrices are crucial for transit agencies to design and operate transit systems. This paper presents a novel temporal Bayesian model designed to estimate transit OD matrices at the individual bus-journey level from boarding/alighting counts at bus stops. Our approach begins by modeling the number of alighting passengers at subsequent bus stops, given a boarding stop, through a multinomial distribution parameterized by alighting probabilities. Given the large scale of the problem, we generate alighting probabilities with a latent variable matrix  and factorize it into a mapping matrix and a temporal matrix, thereby substantially reducing the number of parameters. To further encode a temporally-smooth structure in the parameters, we impose a Gaussian process prior on the columns of the temporal factor matrix. For model inference, we develop a two-stage algorithm with the Markov chain Monte Carlo (MCMC) method. In the first stage, latent OD matrices are sampled conditional on model parameters using a Metropolis-Hastings sampling algorithm with a Markov model-based proposal distribution. In the second stage, we sample model parameters conditional on latent OD matrices using slice and elliptical slice sampling algorithms. We assess the proposed model using real-world data collected from three bus routes with varying numbers of stops, and the results demonstrate that our model achieves accurate posterior mean estimation and outperforms the widely used iterative proportional fitting (IPF) method. Additionally, our model can provide uncertainty quantification for the OD demand matrices, thus benefiting many downstream planning/operational tasks that require robust decisions.
\end{abstract}


\section{Introduction}\label{sec:intro}
The origin-destination (OD) matrix for a bus route captures passenger flows from one stop to another, serving as a comprehensive representation of passenger demand. These matrices can be defined either in an aggregated manner (e.g., overall demand during morning rush hours) or in detail for each bus journey. An accurate estimation of the OD matrices and a good understanding of how these matrices evolve over time are critical for transit agencies in making planning and operational decisions, such as route design \citep{ahern2022approximate}, service scheduling \citep{martinez2014frequency}, timetabling \citep{sun2014demand}, and fleet allocation \citep{gkiotsalitis2019cost}.

Estimating OD matrices for transit systems from available operational data has been an important application and a long-standing research question for both practitioners and researchers \citep{hussain2021transit,mohammed2023origin}. A traditional approach to collecting transit OD matrices is through onboard surveys. This approach, as expected, is time-consuming and labor-intensive \citep{agrawal2017comparing}. More importantly, there could be systematic bias when using the OD matrix obtained from a specific bus journey as the base to estimate the overall demand profile due to the inherent randomness in bus operations. For example, the OD matrices of two consecutive bus journeys could differ substantially from each other when bus bunching occurs. The emergence of advanced data collection techniques has offered alternative solutions to collect/estimate OD data. For example, automated fare collection (AFC) systems in cities such as Beijing and Singapore can register both the boarding and alighting transactions of passenger trips  \citep{sun2014models}. For such cases, the AFC data contains the complete information of the OD matrices. However, most cities around the world adopt a ``tap-in-only'' AFC system without recording the alighting information, thus no longer allowing direct inference of OD matrices. Although there exists a large body of literature on inferring passenger alighting stops from ``tap-in-only'' AFC data based on the trip chain continuity assumption \citep[see e.g.,][]{trepanier2007individual, assemi2020improving, hussain2021transit,mohammed2023origin}, these methods are often inaccurate for infrequent transit users \citep{cheng2021probabilistic}. A more prevalent data collection technique is the Automated Passenger Counting (APC) system, which registers the boarding and alighting counts when the bus arrives at a bus stop. APC systems have been widely used by transit agencies and allow one to estimate the number of passengers onboard during the trip in real time. However, because boarding/alighting counts are in fact column/row sums of an OD matrix, we cannot directly infer the OD matrices from APC data.

\begin{figure}[!t]
\centering
\includegraphics[width = 0.95\textwidth]{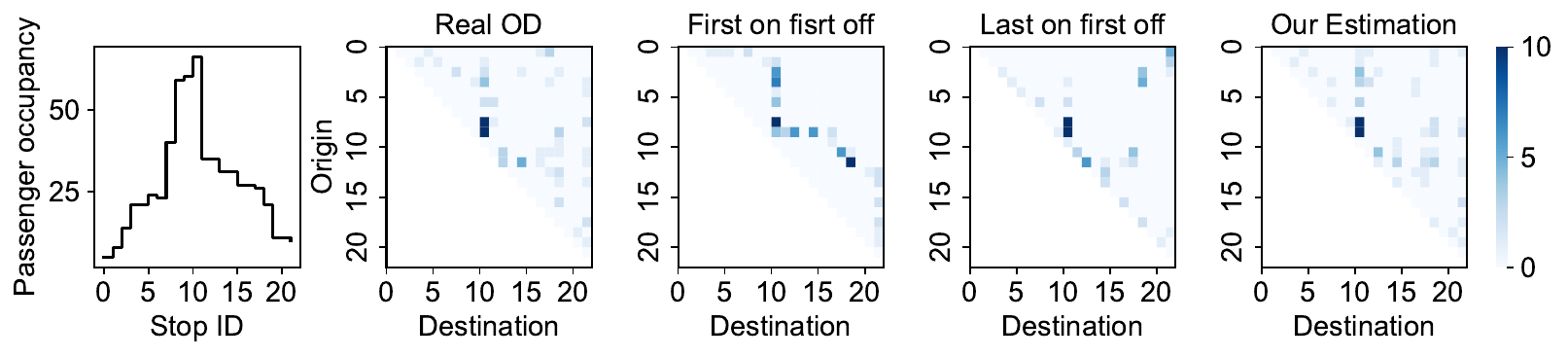}
\caption{The first panel presents the passenger occupancy (y-axis) against the stop number (x-axis). The next four panels present the OD matrices. The first OD matrix is the real OD matrix of a specific bus trip. The last three OD matrices are estimated with different methods. The second matrix is estimated using the ``first on first off'' principle which assumes that passengers who board first will alight first. The third matrix is estimated with the ``last on first off'' principle which assumes that the last passengers to board are the first ones to alight at subsequent stops. The last matrix is estimated from our proposed method and it shows a similar pattern to the real OD matrix.}
\label{fig:uncertainty}
\end{figure}

The focus of our study is to infer time-varying OD matrices from boarding/alighting counts with high temporal resolution (i.e., at the bus-journey level). The challenges in inferring time-varying OD matrices from APC data are summarized as follows: (1) Transit OD flows need to rely on discrete count models and the likelihood structures are more complicated than continuous models. (2) OD matrices are not static, and time-varying models are needed to characterize how demand evolves over time. For example, we expect the OD demand matrices in two consecutive hours to be similar, while the matrix during the morning rush hour should be substantially different from the one during the evening rush hour. When considering individual buses, the variability in OD matrices between adjacent bus journeys can depend on the reliability and uniformity of the operation. In scenarios where services are regular and consistent, the OD matrices observed from two consecutive journeys are likely to be similar. However, in the case of service interruptions or irregularities, such as bus bunching or delays, the OD matrices can differ significantly from one journey to the next. (3) A model that takes entries in the OD matrix as parameters will involve a larger number of parameters and become difficult to estimate, not to mention when estimating time-varying OD matrices. (4) Due to the underdetermined nature of the linear inverse problem, the set of feasible solutions for OD matrices is extraordinarily large. The application of likelihood-based methods for discrete count data in this context would involve an exhaustive enumeration of all possible OD matrices and therefore will face practical computational limitations, especially when considering bus routes with a large number of stops. For a given set of boarding/alighting counts along a bus journey, Figure~\ref{fig:uncertainty} shows the real OD matrix together with potential solutions estimated with different assumptions. We can observe that these solutions could be substantially different from each other. (5) When making operational and planning decisions, we are more interested in the distribution of the underlying OD demand, rather than a point estimate.

While most previous studies have attempted to solve these challenges and have provided various solutions, the time-varying nature of OD matrices in transit systems has not been well explored. Previous studies on OD matrix estimation in transit systems have focused mainly on static models \citep[e.g.,][]{moshe1037alternative,zhao2007estimating,li2009markov,hazelton2010statistical,blume2022bayesian}. These models estimate OD matrices based on data collected over a fixed period, assuming that passenger demand and travel patterns remain constant during that time. While effective for capturing long-term average travel behaviors, static models fail to account for the dynamic nature of passenger demand that occur throughout the day. Existing dynamic OD matrix estimation methods \citep[e.g.][]{hazelton2008statistical,cho2009time,pitombeira2020dynamic} have been primarily designed for road traffic rather than transit systems. These studies typically rely on Kalman filter-based models, where OD flows in road traffic are assumed to follow Gaussian distributions. However, this assumption does not hold for bus-journey OD flows, which exhibit different statistical properties, such as discreteness, over-dispersion, zero inflation, and non-Gaussian distribution. As a result, existing dynamic OD matrix estimation methods developed for road traffic are not directly applicable to transit systems. Therefore, there is a need to develop a tailored estimation method specifically designed to infer time-varying OD matrices in transit systems.

In this paper, we develop a temporal Bayesian model for inferring transit OD matrices at the individual bus level. To model the discrete count data, we
assume that the number of alighting passengers at subsequent bus stops, given a boarding stop, follows a
multinomial distribution. To better characterize the temporal patterns in passenger demand, we assume that the parameters (i.e., assignment probabilities) of the multinomial distribution vary smoothly over time, thus allowing for building a time-varying model using counts observed from a limited number of bus journeys. We introduce a latent variable matrix and use it to parameterize the time-varying multinomial distributions through the softmax transformation. In addition, we propose using matrix factorization to parameterize the latent matrix as the product of a mapping factor matrix and a temporal factor matrix, which substantially reduces the number of parameters. To encode a temporally smooth structure in the matrix, we impose Gaussian process priors on the columns of the temporal factor matrix, which consequently ensure that the assignment probabilities vary smoothly over time. For model inference, we develop a two-stage algorithm based on MCMC. In the first stage, we sample latent OD matrices conditional on parameters using the Metropolis-Hastings sampling algorithm with the proposal distribution proposed by \citet{hazelton2010statistical}, which efficiently bypasses the need to enumerate the large number of feasible OD matrices that align with observed boarding and alighting counts for each bus trip. In the second stage, we sample model parameters conditional on latent OD matrices obtained from the first stage. The key challenge in this step is to efficiently sample latent Gaussian processes with non-Gaussian likelihood, where the posterior no longer has an analytical formulation. To address this issue, we use elliptical slice sampling (ESS), an efficient algorithm developed by \citet{murray2010elliptical}, to sample the temporal factor matrix. We evaluate our proposed model using real-world APC data and true OD matrices from three bus routes in an anonymous city. We compare the performance of the proposed temporal model to a non-temporal variant, and the results show that the temporal Bayesian model outperforms the non-temporal variant, confirming the importance and value of developing a time-varying model. In addition, we also compare our model with the widely used (iterative proportional fitting) IPF method, and the results show that our model can achieve superior performance in deterministic estimation.

The contributions of this work are summarized as follows. First, the proposed OD matrix estimation method focuses on inferring OD matrices for individual bus journeys using observed boarding and alighting counts at each stop. This approach fully leverages the high-resolution data available from bus operations, offering a more detailed and accurate depiction of passenger flows for each journey. We emphasize that OD matrix estimation at the bus-journey level is particularly significant for public transit systems. This perspective not only captures the intricacies of individual journeys but also maximizes the utility of detailed data collected from bus operations, leading to more precise and actionable insights for transit planning and management. Second, this work introduces the novel method for estimating dynamic transit OD matrices, addressing a critical gap in the literature where existing dynamic OD matrix estimation methods developed for road traffic are not applicable to transit systems. This method is specifically designed to capture the unique characteristics and challenges of transit data, such as discrete passenger counts and the temporal variability inherent in bus journeys.
Third, this work makes a theoretical contribution by modeling each row of the OD matrix as a multinomial distribution with alighting probabilities as parameters. To tackle the challenges of temporal dynamics and high-dimensional estimation, this model designs a softmax parameterization combined with matrix factorization, where the temporal factors follow a Gaussian process prior. This approach reduces dimensionality, ensures smooth time-varying parameters, and enhances the model's effectiveness in handling limited data scenarios. Additionally, we propose to use ESS to efficiently sample the latent Gaussian processes with non-Gaussian likelihood.

The remainder of this paper is organized as follows.
In Section~\ref{sec:problem}, we define the problem and introduce the notation used throughout the paper. In Section~\ref{Sec:Method}, we introduce the proposed temporal Bayesian model. We elaborate on the theoretical underpinnings of our approach and explain how it addresses those identified challenges. Section~\ref{sec:inference} develops an efficient inference algorithm based on MCMC, in which elliptical slice sampling is used to sample the temporal factor matrix. We then evaluate the effectiveness and performance of our proposed model based on real-world data in Section~\ref{sec:exp}. Finally, Section~\ref{sec:con} summarizes our key findings and discusses future research directions.

\section{Literature Review}
OD matrix estimation approaches are generally distinguished into road traffic and transit OD estimation, and further categorized into static and dynamic estimation methods. While our work focuses on transit OD matrix estimation, it is essential to consider the relevant literature on OD matrix estimation for road traffic as well. Understanding the OD estimation approaches in road traffic can provide valuable insights and potential techniques applicable to transit systems. Therefore, we review OD matrix estimation from two key perspectives: static estimation and dynamic estimation. By analyzing both static and dynamic approaches in road traffic and transit systems, we aim to highlight the research gaps in transit OD matrix estimation.
\subsection{Static OD Matrix Inference}
Static OD matrix inference involves estimating the OD matrix for a road or transit network under the assumption that travel patterns remain constant over a specific period. In the OD matrix estimation for a transit network, the iterative proportional fitting (IPF) method has long been the primary solution adopted by transit agencies to estimate OD matrices from APC data. IPF estimates OD matrices by adjusting a reference/seed matrix derived from additional data sources such as onboard surveys \citep{moshe1037alternative,ji2014estimating}. Using the observed counts (i.e., row/column sums), the OD matrix is adjusted through an iterative process: For each row in the matrix, each entry is multiplied by a constant, ensuring that the sum of the row is in accordance with the actual count. Then, this adjustment is applied to each column. This process is repeated iteratively for both rows and columns until the matrix entries converge \citep{lamond1981bregman}. There are several major issues associated with the IPF method. First, the accuracy of IPF is highly dependent on the quality of the seed matrix, while in practice creating a good seed matrix is not an easy task---it has the same number of variables as in the OD matrix. In general, these seed matrices come from onboard surveys. However, as mentioned above, the seed matrix obtained could be biased to represent and reproduce the true demand patterns of OD, and thus the quality of the seed matrix becomes very important for IPF. Second, IPF often struggles with the issue of zeros in the seed matrix \citep{moshe1037alternative}, which will remain zero in all iterations. In cases where the reference matrix contains an entire row or column of zeros but the corresponding actual boarding or alighting counts are not zeros, IPF will fail to find a feasible solution. Third, from a statistical perspective, the estimated matrix after convergence is no longer a ``count'' matrix, thus violating the underlying ``count'' nature of passenger demand. This solution is acceptable when we have a large number of passengers; however, at the journey level, the true OD matrix is likely to be a sparse count matrix, and modeling it using continuous values will result in biased estimation.

Bayesian statistical solutions are developed to combine prior information with observed boarding and alighting counts to achieve accurate inference \citep{li2009markov,hazelton2010statistical,blume2022bayesian}. In contrast to deterministic IPF, Bayesian statistical models have the capability to generate posterior distributions for the elements in the OD matrix, thus providing estimates along with their associated uncertainties. In general, these models treat OD matrix estimation as a linear inverse problem and focus on estimating the alighting probabilities instead of the counts \citep[see e.g.,][]{li2007generalized,ji2015transit}. A landmark work in probabilistic inference is the Bayesian approach proposed by \citet{li2009markov}, which incorporates a Markov model to describe the relationships between the entries of a transit OD matrix. This approach assumes that the probability of an onboard passenger alighting at the next approaching stop is independent of his/her boarding stop. In other words, the onboard passengers are assumed to be memoryless. While this innovative approach reduces the model parameters significantly and enables elegant likelihood construction, the Markov assumption is too restrictive and unrealistic for practical applications. To address this limitation, \citet{hazelton2010statistical} introduced a novel Bayesian method for static OD matrix inference. A key challenge in building the statistical model is that the calculation of the likelihood of observed counts requires an enumeration of all possible OD matrices (with elements being counts) that match the marginal counts. To tackle this challenge, \citet{hazelton2010statistical} developed a two-stage sampling algorithm for model inference using the MCMC method. The first stage samples latent OD matrices using the Markov model by \citet{li2009markov} as the proposal distribution. The second stage samples model parameters conditional on the OD matrices in the first stage. \citet{blume2022bayesian} developed a Bayesian inference approach to estimate the static OD matrix in large-scale networked transit systems but considering elements as continuous random variables. This problem is approached as an inverse linear regression, and the posterior distributions of OD matrix entries are estimated using Hamiltonian Monte Carlo. Overall, the statistical methods summarized above are essentially designed for static OD inference, and they generally require many observed bus journeys to estimate one single OD matrix.

\subsection{Dynamic OD Matrix Inference}
Dynamic OD matrix inference involves estimating OD matrices that change over time, capturing the variations in travel patterns throughout the day or over longer periods. Dynamic OD estimation has been extensively studied in road traffic, but it has received less attention in transit systems. We here review the methodologies and advancements in dynamic OD estimation within the context of road traffic.

Dynamic OD estimation methods in road traffic can generally be classified into two main categories: optimization-based models, which are more suited for offline estimation, and Kalman filter-based models, which are applicable for both offline and online estimation \citep{pitombeira2020dynamic}. \cite{cremer1987new} introduced pioneering approaches for estimating dynamic OD flow proportions in traffic facilities, utilizing both optimization and Kalman filter models. Their methods were designed to estimate OD flows in scenarios where total entering and exiting flows are observed, providing a foundation for dynamic OD estimation in traffic systems. Building on this work, \cite{cascetta1993dynamic} advanced the field by extending static optimization models to handle dynamic conditions. In their approach, assignment matrices, which determine how traffic flows are distributed across the network, are estimated exogenously, allowing for more accurate modeling of time-varying traffic patterns. \cite{chang1994recursive} developed the dynamic OD flow estimation model using an extended Kalman filter specifically designed for freeway corridors, providing a method to adapt OD estimates in real-time as new data becomes available. \cite{sherali2001estimation} introduced a parametric optimization approach to estimate time-dependent trip tables or path flows based on link traffic volumes for a road network. Their model aims to find path flows that strike a balance between the least-cost OD paths and those that best match the observed link flows. This compromise allows for a more accurate estimation of time-dependent traffic patterns by integrating both cost-efficiency and traffic data into the estimation process. \cite{ashok2002estimation} introduced a novel model for estimating and predicting time-dependent OD matrices, with a key contribution being the explicit modeling of the dynamic assignment matrix that connects OD flows to link volumes. This approach systematically accounts for the uncertainty in travel times and route choice fractions, which are critical components of the assignment matrix. The empirical results suggest that using the proposed stochastic assignment matrix can yield more accurate results compared to conventional models that rely on a fixed assignment matrix. \cite{zhou2007structural} developed a dynamic OD estimation and prediction model that is integral to real-time dynamic traffic assignment systems. The model utilizes a structural state space approach to incorporate regular demand patterns, structural deviations, and random fluctuations, employing a polynomial trend filter to capture real-time demand changes. By integrating a Kalman filtering framework, the model adaptively updates the regular demand pattern estimates using new daily observations. \cite{hazelton2008statistical} proposed to estimate a sequence of OD matrices from daily link count data by recommending a parsimonious parameterization to facilitate the application of standard statistical estimation techniques. The study introduces a multivariate Gaussian model for the link counts, grounded in an overdispersed Poisson process, and emphasizes the use of Bayesian methods to incorporate prior information effectively. \cite{cho2009time} addressed the limitations of previous time-dependent OD estimation methods that typically assume prior knowledge of the OD matrix by developing a novel approach combining Gibbs sampling and Kalman filtering within a state space model. The proposed solution algorithm incorporates parallel chain convergence control and demonstrates improved efficiency and speed through a parallel computing structure. \cite{xie2015excess} formulated the dynamic OD demand estimation problem as an excess-demand dynamic traffic assignment (DTA) problem within an expanded network using dummy paths. This innovative approach allows for the application of existing DTA solution methods and software tools to derive path flow patterns in the expanded network, which in turn provides the OD demand patterns in the original network. \cite{hazelton2016statistical} addressed the estimation of dynamic OD flows by incorporating a learning mechanism in route choices. They assumed that route flows evolve according to a Markovian transition kernel and that route choices are governed by a logit model, where the utility of each route is a function of past route costs. They proposed the corresponding MCMC method to estimate
OD flows and route choice parameters. \citet{pitombeira2020dynamic} proposed a dynamic hierarchical Bayesian model for estimating day-to-day OD flows in transportation networks, leveraging traffic link volume data collected over several days. The model is structured in three levels: at the first level, a dynamic Gaussian model captures the temporal evolution of OD flows; at the second level, the assignment of route flows is modeled based on OD flows, with route choice probabilities influenced by user-predicted route costs that depend on past experiences and a learning parameter controlling the memory length of users. Finally, at the third level, the model describes the observed link volumes as a function of the route flows, with the covariance matrices of OD flows and link volumes modeled through variance functions.

There are some research gaps in the current literature on transit OD matrix estimation. Firstly, while existing static models can indeed use multi-day count data to estimate OD matrices for specific periods within a day, such as morning peak hours, by pooling data across multiple days to increase sample size and improve estimation accuracy, these static models assume that the parameters within each period are static, which may not hold true in practical scenarios where passenger demand and travel patterns should vary within these periods. This limitation highlights the need for more flexible models that can capture the inherent variability within such periods. Secondly, dynamic OD matrix inference has been extensively studied in the context of road traffic, with much less focus on transit systems. In road traffic, many studies rely on Kalman filter-based models, which are well-suited to real-time estimation and adjustment of OD matrices as new data becomes available. These models often assume that OD flows follow a Gaussian distribution, which is appropriate given the large volumes of traffic data and the continuous nature of vehicle flow on roads. However, the assumption of Gaussian distribution does not hold in this context, as bus-journey data typically exhibit different statistical properties, such as over-dispersion and zero inflation, which are common in count data. Consequently, applying road traffic-based dynamic models to transit systems can lead to inaccurate estimations and an inability to fully capture the variability and dynamics of passenger demand in public transit. This gap underscores the need for specialized dynamic OD inference models tailored to the unique characteristics of transit systems.

\section{Problem Definition}\label{sec:problem} 

Consider a bus route comprising $S$ stops at which passengers can board and alight. Let $u_i$ and $v_i$ denote the numbers of boarding passengers and alighting passenger at stop $i$, respectively, for $i=1,2,\ldots,S$. Such boarding/alighting counts are available from the APC systems. In general, we will see neither alighting passengers at stop 1 nor boarding passengers at stop $S$, so we can fix $v_1=u_S=0$.

\begin{figure}[!t]
\centering
\includegraphics[width = 0.95\textwidth]{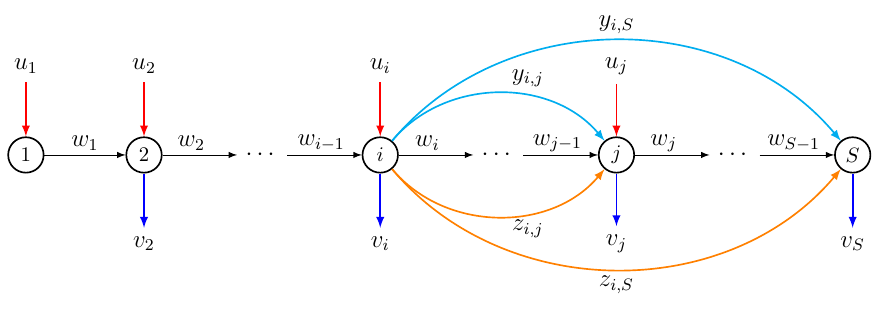}
\caption{A graphical representation of notations in a bus route. There is a bus route with $S$ stops. $u_i$ and $v_i$ are boarding and alighting counts of passengers at stop $i$, respectively; $y_{i,j}$ is the number of passengers boarding at stop $i$ and alighting at stop $j$; $w_i$ is the number of passengers on the bus immediately after leaving stop $i$; $z_{i,j}$ is the number of passengers boarded at stop $i$ and are on the bus as it approaches stop $j$.}
\label{fig:bus_route}
\end{figure}

For a bus journey (i.e., a trip from stop 1 to stop $S$), we denote by $y_{i,j}$ the number of passengers who board at stop $i$ and alight at stop $j$, which cannot be observed directly. We define $w_i$ as the number of passengers on the bus immediately after leaving stop $i$. This can be expressed recursively as
\begin{equation}
w_i = w_{i-1} + u_i - v_i \quad\quad \left(i=1,2,\ldots,S\right),
\end{equation}
with the initial condition $w_0=0$. Let $z_{i,j}$ represent the unobserved number of passengers who board at stop $i$ and remain on the bus as it approaches stop $j$. The relationships among these variables are given by
\begin{align}
z_{j,j+1} &= u_j, \label{eq:zu}\\
z_{i,j+1} &= z_{i,j} - y_{i,j} \quad\quad \left(i = 1,2,\ldots,j-1\right),\label{eq:zf}\\
w_j &= \sum_{i=1}^j z_{i,j}.
\end{align}
These notations of a bus route are graphically represented in Fig.~\ref{fig:bus_route}. Let $\boldsymbol{u} = \left(u_1,u_2,\ldots,u_S\right)^\top$ and $\boldsymbol{v} = \left(v_1,v_2,\ldots,v_S\right)^\top$ be the vectors of boarding and alighting counts at the stops, respectively; we then denote by $\boldsymbol{x} = \left(\boldsymbol{u}^\top,\boldsymbol{v}^\top\right)^\top = \left(u_1,u_2,\ldots,u_S,v_1,v_2,\ldots,v_S\right)^\top$ as the aggregation of observed counts for a bus trip.

Our study aims to infer the OD matrix $\boldsymbol{Y}=\left(y_{i,j}\right)_{S\times S}$. For a bus route/service, it is clear that passengers can only travel to downstream stops. Thus, we fix $y_{i,j}=0$ for all cases where $i\geq j$, and focus exclusively on the upper-triangular part of $\boldsymbol{Y}$.
We denote by $\boldsymbol{y}_i$ the number of passengers traveling from the
$i$-th stop to the subsequent stops along the bus route. For instance, $\boldsymbol{y}_1 = \left(y_{1,2}, y_{1,3}, \ldots, y_{1,S}\right)^\top$ denotes the passenger counts from the initial stop to all subsequent stops along the route. This definition continues to $\boldsymbol{y}_{S-1}=\left(y_{S-1,S}\right)$, which represents the number of passengers traveling from the second-to-last stop to the last stop. Next, we stack these passenger counts into a single OD vector $\boldsymbol{y} = \left({\boldsymbol{y}_1^\top}, {\boldsymbol{y}_2^\top}, \ldots, {\boldsymbol{y}_{S-1}^\top}\right)^\top\in
\mathbb{R}^{S\left(S-1\right)/2}$. 
Although $\boldsymbol{y}$ is not directly observable, its relationship with the observed boarding and alighting counts can be expressed as follows:
\begin{equation}
    \sum_{j=i+1}^Sy_{i,j}=u_i \quad\quad \left(i=1,2,\ldots,S-1\right),
\end{equation}
and
\begin{equation}
    \sum_{i=1}^{j-1}y_{i,j}=v_j \quad\quad \left(j=2,\ldots,S-1\right).
\end{equation}
This relationship can be encapsulated as
\begin{equation}
   \boldsymbol{x} = \textbf{A}\boldsymbol{y}, \label{eq:linear}
\end{equation}
 where both $\boldsymbol{x}$ and $\boldsymbol{y}$ are count-valued vectors, and $\textbf{A}$ is a $2S \times M$ binary routing matrix defined by
\begin{equation}
    a_{i,j} =
\begin{cases}
1, & \text{if } i = 1, \ldots, S \text{ and }\\&j = S\left(i - 1\right) - i\left(i + 1\right)/2 + k \\ &\text{for } k = i + 1, \ldots, S, \\
1, & \text{if } i = S + 1, \ldots, 2S \text{ and }\\&j = S\left(k - 2\right) - k\left(k + 1\right)/2 + i \\ &\text{for } k = 1, \ldots, i - S - 1, \\
0, & \text{otherwise.}
\end{cases}
\end{equation}
Notably, the $S$-th and $\left(S+1\right)$-th rows of $\textbf{A}$ contain only zero elements, corresponding to the non-existent boarding and alighting counts at the terminal and initial stops, respectively. Although these two rows of $\textbf{A}$ are redundant, they can maintain a direct correspondence between the row indices of the matrix and the bus stop numbers. Using a bus route with six stops as an example, the linear relationship between the observation $\boldsymbol{x}$ and the OD vector $\boldsymbol{y}$ can be expressed as

\begin{equation}\label{eq:linear constraints}
    \begin{matrix}
    \underbrace{\begin{bmatrix}
          u_{1}\\
          u_{2}\\
          u_{3}\\
          u_{4}\\
          u_{5}\\
          u_{6}\\
          v_{1}\\
          v_{2}\\
          v_{3}\\
          v_{4}\\
          v_{5}\\
          v_{6}\\
\end{bmatrix}}_{\boldsymbol{x}} = \addtocounter{MaxMatrixCols}{13}
    \underbrace{
        \begin{bmatrix}
{\color{blue}1} & {\color{blue}1} & {\color{blue}1} & {\color{blue}1} & {\color{blue}1} & 0 & 0 & 0 & 0 & 0 & 0 & 0 & 0 & 0 & 0 \\
0 & 0 & 0 & 0 & 0 & {\color{blue}1} & {\color{blue}1} & {\color{blue}1} & {\color{blue}1} & 0 & 0 & 0 & 0 & 0 & 0\\
0 & 0 & 0 & 0 & 0 & 0 & 0 & 0 & 0 & {\color{blue}1} & {\color{blue}1} & {\color{blue}1} & 0 & 0 & 0\\
0 & 0 & 0 & 0 & 0 & 0 & 0 & 0 & 0 & 0 & 0 & 0 & {\color{blue}1} & {\color{blue}1} & 0\\
0 & 0 & 0 & 0 & 0 & 0 & 0 & 0 & 0 & 0 & 0 & 0 & 0 & 0 & {\color{blue}1}\\
0 & 0 & 0 & 0 & 0 & 0 & 0 & 0 & 0 & 0 & 0 & 0 & 0 & 0 & 0\\
0 & 0 & 0 & 0 & 0 & 0 & 0 & 0 & 0 & 0 & 0 & 0 & 0 & 0 & 0\\
{\color{blue}1} & 0 & 0 & 0 & 0 & 0 & 0 & 0 & 0 & 0 & 0 & 0 & 0 & 0 & 0\\
0 & {\color{blue}1} & 0 & 0 & 0 & {\color{blue}1} & 0 & 0 & 0 & 0 & 0 & 0 & 0 & 0 & 0\\
0 & 0 & {\color{blue}1} & 0 & 0 & 0 & {\color{blue}1} & 0 & 0 & {\color{blue}1} & 0 & 0 & 0 & 0 & 0\\
0 & 0 & 0 & {\color{blue}1} & 0 & 0 & 0 & {\color{blue}1} & 0 & 0 & {\color{blue}1} & 0 & {\color{blue}1} & 0 & 0\\
0 & 0 & 0 & 0 & {\color{blue}1} & 0 & 0 & 0 & {\color{blue}1} & 0 & 0 & {\color{blue}1} & 0 & {\color{blue}1} & {\color{blue}1}\\
    \end{bmatrix}}_{\mathbf{A}}\underbrace{\begin{bmatrix}
          y_{1,2}\\
          y_{1,3}\\
          y_{1,4}\\
          y_{1,5}\\
          y_{1,6}\\
          y_{2,3}\\
          y_{2,4}\\
          y_{2,5}\\
          y_{2,6}\\
          y_{3,4}\\
          y_{3,5}\\
          y_{3,6}\\
          y_{4,5}\\
          y_{4,6}\\
          y_{5,6}\\
\end{bmatrix}}_{\boldsymbol{y}}
    \end{matrix}.
\end{equation}

To account for multiple bus journeys, we extend the notation to include a bus index $n$, which represents the $n$-th bus journey, with $\boldsymbol{x}^n=\left({\boldsymbol{u}^n}^\top,{\boldsymbol{v}^n}^\top\right)^\top$ and $\boldsymbol{y}^n = \left({\boldsymbol{y}_1^n}^\top,\ldots,{\boldsymbol{y}_{S-1}^n}^\top\right)^\top$. To effectively model the dynamic/time-varying nature of OD matrices/vectors, our model incorporates temporal information. Specifically, we denote by $t^n$ the departure time at the initial stop for the $n$-th bus trip/journey. For a total of $N$ bus journeys over a studied period, we define $\mathcal{X}=\left\{\boldsymbol{x}^n\mid n=1,2,\ldots,N\right\}$ as the set of observed boarding and alighting counts and $\boldsymbol{t}=\left(t^1,t^2,\ldots,t^N\right)^\top$ as the vector of observed departure time. The primary objective is to estimate the set of OD vectors, denoted as $\mathcal{Y}=\left\{\boldsymbol{y}^n\mid n=1,2,\ldots,N\right\}$, using observed data set $\mathcal{X}$ and $\boldsymbol{t}$. This problem is challenging because the number of unknown quantities (OD vector) is much larger than the number of observations (boarding and alighting counts) in the linear system expressed by Eq.~\eqref{eq:linear}, resulting in a challenging statistical linear inverse problem \citep{vardi1996network,hazelton2010statistical}. We denote by $\mathcal{H}\left(\boldsymbol{x}^n\right) = \left\{\boldsymbol{y}^n\mid \boldsymbol{x}^n = \textbf{A}\boldsymbol{y}^n\right\}$ the solution space that encompasses all feasible OD vectors consistent with the observation $\boldsymbol{x}^n$. In general, the solution space could be very large even for a route with a modest number of stops.

\section{Bayesian Model Formulation}
\label{Sec:Method}

\subsection{Likelihood for Static and Time-varying Models}

\citet{hazelton2010statistical} models the entries in the OD matrix using Poisson distributions and introduces a set of intensity parameters. However, the exact OD entries in two consecutive matrices (e.g., $y_{i,j}^n$ and $y_{i,j}^{n+1}$) could vary substantially due to operational randomness and factors such as bus bunching, and it becomes problematic to use a fixed Poisson intensity to model both observations. On the other hand, we can safely assume that the alighting probabilities are universal for both journeys, and the variation in $y_{i,j}$ is due to the variation in the boarding counts $u_i$. In this paper, we therefore consider each row in the OD matrix as a realization from a multinomial distribution and treat the alighting probabilities for each bus stop as model parameters.

In terms of notation, let $\lambda_{i,j}^n$ be the probability that a passenger boarding at stop $i$ will alight at stop $j$ during the
$n$-th bus trip. Furthermore, let $\boldsymbol{\lambda}_i^n=\left({\lambda_{i,i+1}^n},\ldots,{\lambda_{i,S}^n}\right)^\top$be the alighting probabilities of downstream stops for a passenger boarding at stop $i$, and the sum of these probabilities is one, i.e., $\sum_{j=i+1}^{S} \lambda_{i,j}^n=1$. Next, let $\boldsymbol{\lambda}^n=\left({\boldsymbol{\lambda}^n_1}^\top,\ldots,{\boldsymbol{\lambda}^n_{S-1}}^\top\right)^\top$ denote probabilities for all the corresponding OD entries of the $n$-th bus trip. Assuming that passengers make decisions independently, $\boldsymbol{y}_i^n$ follows a multinomial distribution
$\boldsymbol{y}_i^n\sim \text{Multinomial}\left(u_i^n,\boldsymbol{\lambda}_i^n\right)$. Specifically, it can be represented as
\begin{align} 
    p\left(\boldsymbol{y}_i^n\mid u_i^n,\boldsymbol{\lambda}_i^n\right)=u_i^n!\prod_{j=i+1}^S\frac{{\lambda_{i,j}^n}^{y_{i,j}^n}}{y_{i,j}^n!},\label{eq:mltinomial_likelihood}
\end{align}
and the likelihood of observing $\boldsymbol{x}^n$ becomes
\begin{align}\label{eq:true_like}
L\left(\boldsymbol{\lambda}^n\right)=p\left(\boldsymbol{x}^n\mid\boldsymbol{\lambda}^n\right)&=\sum_{\boldsymbol{y}^n}p\left(\boldsymbol{x}^n\mid\boldsymbol{y}^n,\boldsymbol{\lambda}^n\right)p\left(\boldsymbol{y}^n\mid\boldsymbol{\lambda}^n\right)\notag\\
    &=\sum_{\boldsymbol{y}^n\in\mathcal{H}\left(\boldsymbol{x}^n\right)}p\left(\boldsymbol{y}^n\mid\boldsymbol{\lambda}^n\right)\notag\\
    &=\sum_{\boldsymbol{y}^n\in\mathcal{H}\left(\boldsymbol{x}^n\right)}\prod_{i=1}^{S-1}u_i^n!\prod_{j=i+1}^S\frac{{\lambda_{i,j}^n}^{y_{i,j}^n}}{y_{i,j}^n!}.
\end{align}

Clearly, such a model is not identifiable if we have only one bus journey. \citet{hazelton2010statistical} assumes that over a certain period of time we have access to repeated and independent observations of $\boldsymbol{x}$ following the same distribution with parameters $\boldsymbol{\lambda}$. Under such an assumption, $\boldsymbol{\lambda}$ becomes identifiable. For modeling multiple bus journeys in a day, we expect $\boldsymbol{\lambda}_i^n$ to vary smoothly from one bus to the next (or over time). In this case, we need an effective parameterization that produces time-varying multinomial probabilities. Since $\boldsymbol{\lambda}_i$ is
the parameter of a multinomial distribution, \citet{hazelton2010statistical} in the discussion on non-Poisson boarding process suggested using conjugate Dirichlet priors for $\boldsymbol{\lambda}_i$, which can model the uncertainty about $\boldsymbol{\lambda}_i$. However, in practice, it becomes challenging to adapt the Dirichlet distribution and encode temporal dynamics to generate time-varying samples of $\{\boldsymbol{\lambda}_i^n\}$.

\subsection{Parametrization of Time-varying Assignment Probabilities}

As mentioned, although the Dirichlet distribution is a natural prior for modeling $\boldsymbol{\lambda}_i$, it is difficult to adapt it to dynamic/time-varying processes. To address this issue and effectively characterize the time-varying nature of $\boldsymbol{\lambda}_i^n$, we employ a natural softmax parameterization:
\begin{equation}
    \begin{aligned}
\boldsymbol{\lambda}_i^n=\text{Softmax}\left(\rho\boldsymbol{G}_i^n\right) =  \begin{pmatrix}
 \frac{\exp\left(\rho G_{i,i+1}^n\right)}{1+\sum_{j=i+1}^{S-1}\exp\left(\rho G_{i,j}^n\right)}\\
 \frac{\exp\left(\rho G_{i,i+2}^n\right)}{1+\sum_{j=i+1}^{S-1}\exp\left(\rho G_{i,j}^n\right)}
 \\
 \vdots
 \\
 \frac{\exp\left(\rho G_{i,S-1}^n\right)}{1+\sum_{j=i+1}^{S-1}\exp\left(\rho G_{i,j}^n\right)} \\
  \frac{1}{1+\sum_{j=i+1}^{S-1}\exp\left(\rho G_{i,j}^n\right)}
\end{pmatrix},\label{eq:softmax}
    \end{aligned}
\end{equation}
where $\boldsymbol{G}_i^n = \left(G_{i,i+1}^n,G_{i,i+2}^n,\ldots,G_{i,S-1}^n\right)^\top \in \mathbb{R}^{S-i-1}$, and $\rho>0$ is the temperature parameter, which can help to learn good sharpness/smoothness of the probability distribution. Next, we denote the collection of $\boldsymbol{G}_i^n$ over $N$ bus journeys by the matrix
\begin{equation}
    \boldsymbol{G}_i = \left[\boldsymbol{G}_i^1,\boldsymbol{G}_i^2,\ldots,\boldsymbol{G}_i^N\right].
\end{equation}

Now we can see that $\boldsymbol{G}_i$ contains $(S-i-1)\times N$ parameters to be estimated. It should also be noted that there is no need to introduce $\boldsymbol{G}_{S-1}$ as there is only one possible alighting stop, i.e., stop $S$, and we always have $\lambda_{S-1,S}^n=1$. For dynamic models, a general approach in the literature is to impose a state-space model with Gaussian noise, for instance:
\begin{equation} \label{eq:dtm}
    \boldsymbol{G}_i^n \mid  \boldsymbol{G}_i^{n-1} \sim \mathcal{N}\left(\boldsymbol{G}_i^{n-1},\sigma^2\boldsymbol{I}\right).
\end{equation}
This parameterization of time-varying multinomial probabilities has been used in dynamic topic models \citep{blei2006dynamic}. However, for the entire bus route, we have to create a dynamic model for each bus stop $i=1,\ldots,S-2$, and the number of variables in the latent state for each bus becomes $\left(S-1\right)\left(S-2\right)/2$. Despite having a simple formulation, this parameterization shows several critical issues. First, we have a non-Gaussian state-space model in which the likelihood of observing $\boldsymbol{x}^n$ is computationally intractable (see Eq.~\eqref{eq:true_like}). Although the likelihood of $\boldsymbol{y}^n$ can be computed, estimating $\boldsymbol{G}_i$ requires filtering based on the non-Gaussian likelihood, which becomes computationally prohibitive given the large dimensionality of the latent state. Second, the state transition model in Eq.~\eqref{eq:dtm} assumes that the $(S-1)(S-2)/2$ latent states vary independently over time, which ignores potential structures over space and time. For example, in transportation systems, it is likely that the probabilities $\lambda_{i,k}$ and $\lambda_{j,k}$ ($i\neq j$) share similar temporal patterns, which is determined by the land-use profile of stop $k$.

To address these issues, we next introduce an alternative parameterization for $\boldsymbol{G}_i$. Specifially, we assume $\boldsymbol{G}_i$ has a low-rank structure: 
\begin{align}
    \boldsymbol{G}_i = \boldsymbol{\Phi}_i\boldsymbol{\Psi}^\top = \sum_{d=1}^D\boldsymbol{\phi}_{i,d}\boldsymbol{\psi}_d^\top, \label{eq:matf}
\end{align}
where $\boldsymbol{\Phi}_i\in \mathbb{R}^{(S-i-1)\times D}$, $\boldsymbol{\Psi_{d}}\in \mathbb{R}^{N\times D}$, and $\boldsymbol{\phi}_{i,d}$ and $\boldsymbol{\psi}_{d}$ are the $d$-th column of $\boldsymbol{\Phi}_i$ and $\boldsymbol{\Psi}$, respectively. Stacking Eq.~\eqref{eq:matf} for bus stops $i=1,\ldots,S-2$ together, we have
\begin{equation} \label{eq:gpfa}
\boldsymbol{G}=\begin{bmatrix}
\boldsymbol{G}_1\\ 
\vdots\\ 
\boldsymbol{G}_{S-2}
\end{bmatrix} = \begin{bmatrix}
\boldsymbol{\Phi}_1\\ 
\vdots\\ 
\boldsymbol{\Phi}_{S-2}
\end{bmatrix} \times \boldsymbol{\Psi}^\top = \boldsymbol{\Phi}\boldsymbol{\Psi}^\top,
\end{equation}
where we refer to $\boldsymbol{\Phi}$ as the \textit{mapping factor matrix} and $\boldsymbol{\Psi}$ as the \textit{temporal factor matrix}. The low-rank assumption posits that $D\ll N$ and $D\ll (S-2)(S-1)/2$, so that the factorization  
 of $\boldsymbol{G}$ substantially reduces the number of parameters.

In order to encode temporal smoothness in $\boldsymbol{G}$, we assume that each column $\boldsymbol{\psi}_{d}$ in $\boldsymbol{\Psi}$ is generated from a latent Gaussian process by taking values at bus departure times $\boldsymbol{t}$ with kernel/covariance function $k_d\left(t,t'; \boldsymbol{\eta}_d\right)$ where $\boldsymbol{\eta}_d$ is the vector of kernel hyperparameters. For example, a widely used kernel function that can produce smooth functions is the squared-exponential kernel:
\begin{equation}
k\left(t,t';l,\sigma^2\right) = \sigma^2\exp\left(-\frac{(t-t')^2}{2l^2}\right),
\end{equation}
with two hyperparamters---lengthscale $l$ and variance $\sigma^2$. We further assume the mean of the Gaussian process to be zero, and this gives for $d=1,\ldots,D$
\begin{align} \label{eq:columns}
    \boldsymbol{\psi}_{d}\sim \mathcal{N}\left(\boldsymbol{0}_{N},\boldsymbol{K}_d\right),  \ [\boldsymbol{K}_d]_{ij} = k_d(t_i,t_j;\boldsymbol{\eta}_d).
\end{align}

\begin{figure}[!t]
\centering
\includegraphics[width = 0.85\textwidth]{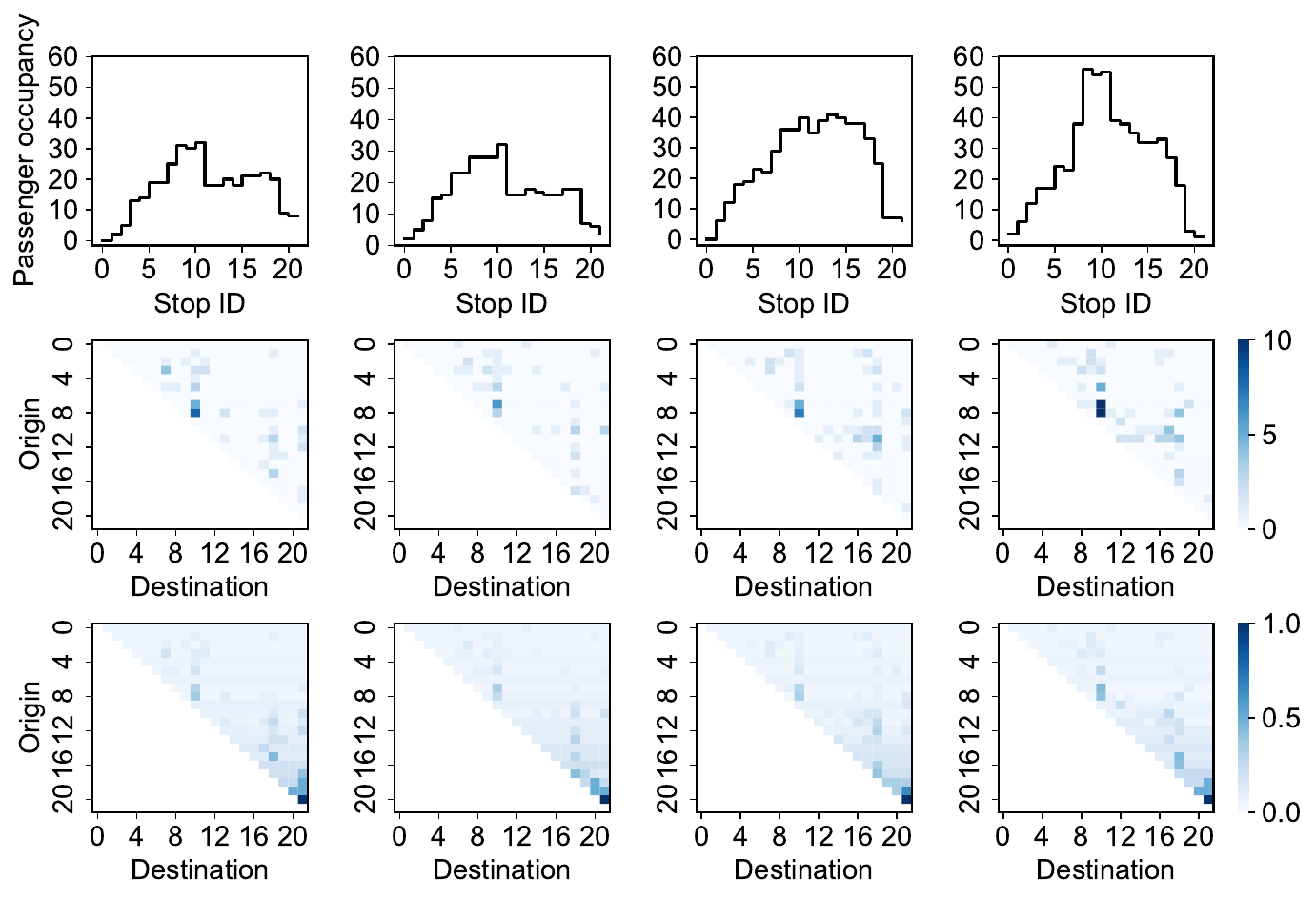}
\caption{Illustration of the smoothly time-varying alighting probabilities. Each row presents the profiles of the four consecutive buses: The first row presents the passenger occupancy (y-axis) against the stop number (x-axis); The second row presents the actual OD matrices; The third row presents the empirical alighting probabilities at stops.}
\label{fig:p}
\end{figure}

The Gaussian process factor model specified in Eqs.~\eqref{eq:gpfa} and \eqref{eq:columns} provides an effective framework to model high-dimensional processes with a temporal structure, and it has been extensively used to model high-dimensional spatial and temporal data \citep[see e.g., in][]{lopes2008spatial,luttinen2009variational,lei2022bayesian}. 
 The specification of $\boldsymbol{\psi}_{d}$ in Eq.~\eqref{eq:columns} facilitates temporally-smooth variations of $G_i$. Consequently, the multinomial probability $\boldsymbol{\lambda}_i$ also exhibits smooth temporal variations. This specification assumes that $\boldsymbol{\lambda}_i^n\approx \boldsymbol{\lambda}_i^{n+1}$, which corresponds to a homogeneous assumption between the two groups of passengers who board bus $n$ and bus $n+1$, respectively, at stop $i$.  This assumption is reasonable, considering that both groups of passengers essentially arrive at stop $i$ at the same time. To illustrate this, we use empirical OD matrices and alighting probabilities from a sequence of four consecutive buses. In Figure~\ref{fig:p}, the first row shows passenger occupancy plotted against the stop number, revealing variations in the number of passengers on different buses. The second row displays the actual OD matrices for each bus, highlighting noticeable differences in passenger flows, even between two consecutive buses. Despite these differences in the OD matrices, the third row, which presents the empirical alighting probabilities for these buses, shows a strong similarity in the alighting patterns. This consistency in alighting probabilities across different buses supports our assumption that while OD demand may fluctuate, the alighting probabilities change smoothly over time. This smooth temporal variation in alighting behavior justifies our modeling approach and indicates that it is a reasonable and effective way to capture the dynamics of transit OD passenger flow.

\subsection{Prior Specification}

In our numerical experiment, for simplicity, we assume that all columns in $\boldsymbol{\Psi}$ are generated independently from the same Gaussian process with a squared-exponential kernel, $\boldsymbol{\eta}_d=\{\sigma,l\}$, with $\sigma=1$ and $l=3600$ sec. We set $\sigma=1$ because $\boldsymbol{\Phi}\boldsymbol{\Psi}^\top = \sigma \boldsymbol{\Phi}\times\frac{1}{\sigma}\boldsymbol{\Psi}^\top$ so it is not necessary to introduce the variance hyperparameter. For lengthscale $l$, we can further put a prior distribution on it to make the model fully Bayesian; however, learning covariance hyperparameters in latent Gaussian process models is known to be a challenging task with convergence issues \citep{murray2010slice}. For model simplicity, we fix $l=1$ hour based on prior knowledge.

For factor matrix $\boldsymbol{\Phi}_i$ ($i=1,\ldots,S-2$), we simply put an independent univariate Gaussian prior for each entry. Alternatively, we have the column $\boldsymbol{\phi}_{i,d}$ to be independent and identically distributed following a zero-mean isotropic Gaussian distribution: 
\begin{align}
    \boldsymbol{\phi}_{i,d} \sim \mathcal{N}\left(\boldsymbol{0}_{S-i-1},\sigma^2_0\boldsymbol{I}\right), \quad d=1,\ldots,D,
\end{align}
where we set hyperparameter $\sigma^2_0=1$.

As the temperature parameter has to be positive, we use Gaussian prior on the log-transformed $\rho$:
\begin{equation}
\log{\left(\rho\right)}\sim \mathcal{N}\left(\mu_{\rho},\sigma^2_{\rho}\right),    
\end{equation}
where we set hyperparameters $\mu_{\rho}=\ln (0.1)$ and $\sigma_{\rho}^2=1$.
 
The overall data generation process can be summarized as: 
\begin{itemize}
\item Draw the temperature parameter $\rho$: $\log{\left(\rho\right)}\sim \mathcal{N}\left(\mu_{\rho},\sigma^2_{\rho}\right)$;
    \item Draw mapping factor matrix $\boldsymbol{\Phi}$: $\boldsymbol{\phi}_{i,d}\sim \mathcal{N}\left(\boldsymbol{0}_{S-i-1},,\sigma^2_0\boldsymbol{I}\right), i=1,\ldots,S-1, d=1,\ldots,D$;
    \item Draw temporal factor matrix $\boldsymbol{\Psi}$: $\boldsymbol{\psi}_d\sim \mathcal{N}\left(\boldsymbol{0},\boldsymbol{K}_d\right), d=1,\ldots,D$;
    \item Obtain parameter matrices $\left\{\boldsymbol{G}_i\right\}_{i=1}^{S-1}$: $\boldsymbol{G}_i = \sum_{d=1}^D\boldsymbol{\phi}_{i,d}\boldsymbol{\psi}_{d}, i=1,\ldots,S-1$;
    \item Obtain parameters $\left\{\boldsymbol{\lambda}_i^n\right\}_{n=1,i=1}^{N,S-1}$: $\boldsymbol{\lambda}_i^n = \text{Softmax}\left(\rho\boldsymbol{G}_i^n\right), n=1,\ldots,N,i=1,\ldots,S-1$;
    \item Draw OD trip vectors $\left\{\boldsymbol{y}^n = \left({\boldsymbol{y}^n_1}^\top,\ldots,{\boldsymbol{y}^n_{S-1}}^\top\right)^\top\right\}_{n=1}^N$:
    
   \hspace{30pt} $\boldsymbol{y}_i^n\sim \text{Multinomial}\left(u_i,\boldsymbol{\lambda}_i^n\right),n=1,\ldots,N,i=1,\ldots,S-1$;
   \item Obtain boarding and alighting observations $\left\{\boldsymbol{x}^n\right\}_{n=1}^N$: $\boldsymbol{x}^n =  \textbf{A}\boldsymbol{y}^n,n=1,\ldots,N$;
\end{itemize} 

\section{Bayesian Inference}\label{sec:inference} 
In this section, we focus on performing Bayesian inference for the proposed model using MCMC. Let $\Theta=\left\{\boldsymbol{\Phi},\boldsymbol{\Psi}, \rho\right\}$ be the set of model parameters. Given $\mathcal{X}$ and $\boldsymbol{t}$, our aim is to infer the posterior distributions of $\Theta$ and $\mathcal{Y}$. The joint posterior of $\Theta$ and $\mathcal{Y}$ can be specified by
\begin{align}    p\left(\Theta,\mathcal{Y}\mid \mathcal{X}, \boldsymbol{t}\right) &= p\left(\Theta\mid
 \mathcal{Y},\boldsymbol{t}\right)p\left(\mathcal{Y}\mid \mathcal{X}\right)\notag\\
 &\propto p\left(\Theta\mid\boldsymbol{t}\right)p\left(\mathcal{Y}\mid \Theta\right)p\left(\mathcal{Y}\mid\mathcal{X}\right)\notag\\
 &\propto p\left(\Theta\mid\boldsymbol{t}\right)\prod_{n=1}^Np\left(\boldsymbol{y}^n\mid \Theta\right)\textit{I}\left(\boldsymbol{y}^n\in \mathcal{H}\left(\boldsymbol{x}^n\right)\right),\label{eq:Post}
\end{align}
where $\textit{I}\left(E\right)$ is the indicator variable for the event $E$. The factorization of the posterior distribution in Eq.~\eqref{eq:Post} naturally provides a two-stage iterative sampling algorithm: (1) draw $\mathcal{Y}$ conditional on $\Theta$ and $\mathcal{X}$, and (2) draw $\Theta$ conditional on $\mathcal{Y}$ and $\boldsymbol{t}$.

There are two critical challenges in the sampling process. First, for the sampling of OD vector $\boldsymbol{y}^n$, computing the likelihood of observing $\boldsymbol{x}^n$ in Eq.~\eqref{eq:true_like} involves the enumeration of all solutions in $\mathcal{H}\left(\boldsymbol{x}^n\right)$, which becomes computationally intractable. Second, columns in the temporal $\boldsymbol{\Psi}$ are in fact latent Gaussian processes with strong dependencies, which require careful consideration when designing the MCMC method.

For the first challenge, \citet{hazelton2010statistical} has provided an effective Metropolis-Hastings sampling solution to generate a candidate OD vector $\boldsymbol{y}^n$ from $\mathcal{H}\left(\boldsymbol{x}^n\right)$ rather than trying to enumerate it. This Metropolis-Hastings sampling strategy can be directly integrated into our time-varying model without much adaptation. We briefly illustrate this approach in Section~\ref{sec:41}.

\subsection{Conditional Sampling of OD Vectors} \label{sec:41}
The subsection gives a brief summary of the method to sample the OD vector $\boldsymbol{y}^n$ conditional on parameters $\Theta$ and boarding/alighting counts $\boldsymbol{x}^n$ developed by \citet{hazelton2010statistical}.
The conditional distribution of $\mathcal{Y}$ is given by
\begin{align}
    p\left(\mathcal{Y}\mid \Theta, \mathcal{X}\right)\propto \prod_{n=1}^Np\left(\boldsymbol{y}^n\mid \Theta\right)\textit{I}\left(\boldsymbol{y}^n\in \mathcal{H}\left(\boldsymbol{x}^n\right)\right).
\end{align}

The Metropolis-Hastings approach for sampling $\boldsymbol{y}^n$ is summarized as follows.
\begin{itemize}
    \item Sample candidate $\boldsymbol{y}^*$ from proposal distribution $q^n$: 
    Let $\boldsymbol{z}_j^*=\left(z_{1,j}^*,\ldots,z_{j-1,j}^*\right)^\top$, where $z_{i,j}^*$ be the candidate number of passengers who boarded at stop
    $i$ and are currently on the bus as it approaches stop $j$. This approach assumes that the passengers on board have the same alighting probability at any stop. Consequently, one can randomly sample $\left(y^*_{1,j},\ldots,y^*_{j-1,j}\right)$ at stop $j$ from $\boldsymbol{z}_j^*$ with the constraint $\sum_{i=1}^{j-1}y^*_{i,j}=v_j$. Then $\boldsymbol{z}^*_{j+1}$ can be updated with Eq.~\eqref{eq:zu} and Eq.~\eqref{eq:zf}, which can be used for the sampling at stop $j+1$. The Markov chain $\left\{\boldsymbol{z}_j^*\right\}$ has transition probability given by
\begin{align}
    \pi_{j+1}=P\left(\boldsymbol{z}_{j+1}^*\mid \boldsymbol{z}_j^*\right) = \left\{\begin{pmatrix}
 w_{j-1}\\v_j
\end{pmatrix}\right\}^{-1}\prod_{i=1}^{j-1}\begin{pmatrix}
 z_{i,j}^*\\y_{i,j}^*
\end{pmatrix}.
\end{align}
Through the process, we can sample $\boldsymbol{y}^*$ and calculate the probability density by
$q^n\left(\boldsymbol{y}^*\right)=\prod_{j=1}^{N-1}\pi_j$.
    \item Update the OD vector:  Accept $\boldsymbol{y}^*$ with probability $\min\left(1,\frac{p\left(\boldsymbol{y}^*\mid \Theta\right)q^n\left(\boldsymbol{y}^n\right)}{p\left(\boldsymbol{y}^n\mid \Theta\right)q^n\left(\boldsymbol{y}^*\right)}\right)$.
If candidate $\boldsymbol{y}^*$ is accepted, then $\boldsymbol{y}^n$ is updated to equal $\boldsymbol{y}^*$. If the candidate is not accepted, $\boldsymbol{y}^n$ remains unchanged.
\end{itemize}

It should be noted that OD vectors for all bus journeys can be sampled in a parallel manner because they are conditionally independent given model parameters and observed counts. The independent proposal assumes that passengers are memoryless, which corresponds to a hypergeometric distribution. While this assumption simplifies the sampling process, it may not be sufficiently accurate when the $\lambda_j$ column exhibits significant heterogeneity. In scenarios where there is considerable variation in $\lambda_j$, the independent proposal might fail to capture the true distribution of passenger behavior effectively. To address this potential limitation, the noncentral hypergeometric distribution could be used as the proposal distribution to account for the weights associated with different alighting probabilities \citep{mccullagh2019generalized,carvalho2014bayesian}. This distribution adjusts the sampling process to reflect the varying probabilities more accurately, thus enhancing the precision of the OD flow estimates. However, the use of noncentral hypergeometric distribution introduces additional complexities in sampling and likelihood calculation. We leave the exploration of this approach to future research.

\subsection{Conditional Sampling of Model Parameters}

The second stage is to sample the model parameters, including the temperature parameter $\rho$, the mapping factor matrix $\boldsymbol{\Phi}$ and the temporal factor matrix $\boldsymbol{\Psi}$, conditional on $\mathcal{Y}$. For these three parameters, a straightforward solution is to use Gibbs sampling to sequentially sample from:
\begin{itemize}
    \item $p\left(\boldsymbol{\Phi}\mid\boldsymbol{\Psi},\rho,\mathcal{Y}\right)$,
    \item $p\left(\boldsymbol{\Psi}\mid\boldsymbol{\Phi},\rho,\mathcal{Y}\right)$,  
    \item $p\left(\rho\mid \boldsymbol{\Psi},\boldsymbol{\Phi},\mathcal{Y}\right)$.
\end{itemize}
However, as mentioned, since the likelihood is multinomial, we can no longer derive the analytical posterior distributions for $\boldsymbol{\Phi}$ and $\boldsymbol{\Psi}$. We next introduce in detail the solution for sampling $\boldsymbol{\Psi}$. 

Given the independent assumption for the columns in $\boldsymbol{\Psi}$, we can sample the whole matrix in a column-based manner. Taking the column $\boldsymbol{\psi}_d$ as an example, we can update $\boldsymbol{\psi}_d$ conditional on $\boldsymbol{\Psi}_{:,h,h\neq d}$, which represents the matrix obtained by removing the $d$-th column vector from $\boldsymbol{\Psi}$. The posterior distribution $p\left(\boldsymbol{\psi}_d\mid \mathcal{Y},\boldsymbol{\Phi},\boldsymbol{\Psi}_{:,h,h\neq d},\rho\right)$ is proportional to the product of the multinomial likelihood $L(\boldsymbol{\psi}_d) = p\left(\mathcal{Y}\mid \boldsymbol{\psi}_d,\boldsymbol{\Psi}_{:,h,h\neq d},\boldsymbol{\Phi},\rho\right)$ and the Gaussian process prior specified by $\mathcal{N}\left(\boldsymbol{\psi}_d ; \boldsymbol{0}_N,\boldsymbol{K}_d\right)$. The problem becomes sampling a latent Gaussian process in a non-conjugate setting. For such problems, \citet{murray2010elliptical} has developed an ESS method that can efficiently explore the parameter space without the need for manually tuning of step sizes or the proposal distributions. This efficiency is achieved by proposing new samples in a manner that directly leverages the underlying correlation structure of the Gaussian process, and ESS has shown superior performance over other methods. Therefore, we use ESS to update $p\left(\boldsymbol{\psi}_d\right)$ and summarize the procedure in Algorithm~\ref{alg:column}. The calculation of likelihood $L(\boldsymbol{\psi}_d)$ is straightforward following Eq.~\eqref{eq:mltinomial_likelihood}:
\begin{align}\label{eq:philike}
     L\left(\boldsymbol{\psi}_d\right)
     &=  p\left(\mathcal{Y}\mid \boldsymbol{\Phi},\boldsymbol{\Psi},\rho\right)= \prod_{n=1}^N p\left(\boldsymbol{y}^n\mid\boldsymbol{\lambda}^n
    \right),
\end{align}
where $\boldsymbol{\lambda}^n$ are computed using the current values of all parameters.

\begin{algorithm}[!t] 
\renewcommand{\algorithmicrequire}{\textbf{Input:}}
\renewcommand{\algorithmicensure}{\textbf{Output:}}
\caption{Elliptical slice sampling for each column $\boldsymbol{\psi}_d$ of factor matrix $\boldsymbol{\Psi}$.}\label{alg:Psi}
\begin{algorithmic}[1]\label{alg:column}
\REQUIRE Current state $\boldsymbol{\psi}_d$, covariance matrix $\boldsymbol{K}_d$, likelihood function $L(\boldsymbol{\psi}_d)$
\ENSURE a new state $\boldsymbol{\psi}_d^\prime$
\STATE Choose ellipse: $\boldsymbol{\nu}\sim \mathcal{N}\left(\boldsymbol{0}_N,\boldsymbol{K}_d\right)$
\STATE Log-likelihood threshold: $\gamma \sim \text{Uniform}\left[0,1\right], \log{c} = \log{L\left(\boldsymbol{\psi}_d\right)} + \log{\gamma}$
\STATE Draw an initial sampling range: $\theta\sim \text{Uniform}\left[0,2\pi\right],\theta_{\min} = \theta-2\pi,\theta_{\max
        }=\theta$
\STATE $\boldsymbol{\psi}_d^\prime= \boldsymbol{\psi}_d\cos{\theta}+\boldsymbol{\nu}\sin{\theta}$
\IF{$\log L\left(\boldsymbol{\psi}_d^\prime\right)>\log c$}
\RETURN $\boldsymbol{\psi}_d^\prime$
\ELSE
\STATE Shrink the sampling range and try a new point:
\STATE \textbf{if} $\theta\leq0$ \textbf{then:} $\theta_{\min}= \theta$ \textbf{else:} $\theta_{\max}=\theta$
\STATE $\theta\sim \text{Uniform}\left[\theta_{\min},\theta_{\max}\right]$
\STATE \textbf{GoTo} Step 4.
\ENDIF
\end{algorithmic}
\end{algorithm}

For sampling $\boldsymbol{\Phi}$, a straightforward approach is to use an element-wise Metropolis-Hasting algorithm given the independent prior. However, as $\boldsymbol{\Phi}$ contains a large number of entries, entry-by-entry sampling is computationally too expensive due to the considerable cost of likelihood evaluation. For efficiency, we sample $\boldsymbol{\Phi}$ in a block-based manner, i.e., updating $\boldsymbol{\Phi}_i$ one by one for $i=1,\ldots,S-2$. For a given block $\boldsymbol{\phi}_{i}$, we can once again use elliptical slice sampling to update each column $\boldsymbol{\phi}_{i,d}$ in $\boldsymbol{\Phi}_i$. The likelihood term can be computed in the same way, and the key difference from the procedure for $\boldsymbol{\psi}_{d}$ is that the prior distribution becomes $\mathcal{N}\left(\boldsymbol{\phi}_{i,d}; \boldsymbol{0}_{S-i-1},\sigma_0^2\boldsymbol{I}\right)$.

For the temperature parameter $\rho$, the posterior distribution is $p\left(\rho\mid\boldsymbol{\Phi},\boldsymbol{\Psi},\mathcal{Y}\right)\propto p\left(\mathcal{Y}\mid\boldsymbol{\Phi},\boldsymbol{\Psi},\rho\right)p\left(\rho\right)$. We propose using slice sampling for $\rho$ and the algorithm is summarized in Algorithm~\ref{alg:rho}. The likelihood $L\left(\rho\right)$ has the same formulation as in Eq.~\eqref{eq:philike}.

\begin{algorithm}[!t]
\renewcommand{\algorithmicrequire}{\textbf{Input:}}
\renewcommand{\algorithmicensure}{\textbf{Output:}}
\caption{Slice sampling for temperature parameter $\rho$.}\label{alg:rho}
\begin{algorithmic}[1]

\REQUIRE Current state $\rho$, likelihood function $L\left(\rho\right)$, slice sampling scale $\epsilon$
\ENSURE a new state $\rho^\prime$
\STATE Log-likelihood threshold: $\gamma\sim \text{Uniform}\left[0,1\right], \log{c}=\log{L\left(\rho\right)}+ \log p(\rho) + \log{\gamma}$
\STATE Draw an initial sampling range:
$\kappa\sim \text{Uniform}\left[0,\epsilon\right]$, $\rho_{\min}=\rho-\kappa$, $\rho_{\max}=\rho_{\min}+\epsilon$
\STATE $\rho^\prime\sim \text{Uniform}\left[\rho_{\min},\rho_{\max}\right]$
\IF{$\log{L\left(\rho^\prime\right)} + \log p(\rho^\prime)>\log{c}$}
\RETURN $\rho^\prime$
\ELSE
\STATE Shrink the sampling range:
\STATE \textbf{if} $\rho^\prime<\rho$ \textbf{then:} $\rho_{\min}= \rho^\prime$ \textbf{else:} $\rho_{\max}=\rho^\prime$
\STATE \textbf{GoTo} Step 3.
\ENDIF
\end{algorithmic}
\end{algorithm}

Finally, we summarize the overall Metropolis-within-Gibbs sampling procedure for estimating the OD vectors and parameters in Algorithm~\ref{alg:overall}.

\begin{algorithm}[!t]
\renewcommand{\algorithmicrequire}{\textbf{Input:}}
\renewcommand{\algorithmicensure}{\textbf{Output:}}
\caption{Metropolis-within-Gibbs sampling for parameters and OD estimation.}\label{alg:overall}
\begin{algorithmic}[1]
\REQUIRE Observations and prior parameters $\mathcal{X}$, $\boldsymbol{t}$, $\sigma$, $l$, $\sigma_0^2$, $\sigma_{\rho}^2$, $\mu_{\rho}$; random initialized factor matrix $\boldsymbol{\Phi}$ and $\boldsymbol{\Psi}$ from matrix normal distribution; the number of burn-in iterations $M_1$, the number of iterations used for estimation $M_2$
\ENSURE Samples of OD vectors $\left\{\mathcal{Y}^{(m)}\right\}_{m=1}^{M_2}$, samples of alighting probabilities $\left\{\boldsymbol{\lambda}^{n(m)}\right\}_{n,m=1}^{N,M_2}$
\FOR{$m=1$ to $M_1+M_2$}
\STATE Update alighting probabilities $\left\{\boldsymbol{\lambda}^{n(m)}\right\}_{n}^{N}$
\STATE Sample OD vectors $\mathcal{Y}^{(m)}$ with Metropolis sampling method
\IF{$m>M_1$}
\RETURN Collect $\mathcal{Y}^{(m)}$ and $\left\{\boldsymbol{\lambda}^{n(m)}\right\}_{n}^{N}$
\ENDIF
\STATE Sample temporal factor matrix $\boldsymbol{\Psi}$ with ESS algorithm 
\STATE Sample mapping factor matrix $\boldsymbol{\Phi}$ with ESS algorithm 
\STATE Sample temperature parameter $\rho$ with slice sampling algorithm
\ENDFOR
\end{algorithmic}
\end{algorithm}

\subsection{Approximating Posterior Distribution of OD Vectors}
In the Bayesian framework, the posterior distribution of an OD vector $\boldsymbol{y}^n$ conditional on observed counts and departure times is obtained by integrating out the model parameters:
\begin{align}
p\left(\boldsymbol{y}^n\mid \mathcal{X},\boldsymbol{t}\right) &= \int p\left(\boldsymbol{y}^n\mid\boldsymbol{x}^n,\Theta\right)p\left(\Theta\mid\mathcal{X},\boldsymbol{t}\right)d\Theta\notag\\ 
&\approx \frac{1}{M}\sum_{m=1}^Mp\left(\boldsymbol{y}^n\mid\boldsymbol{x}^n,\Theta^{\left(m\right)} \right),
\end{align}
where $M$ is the number of samples used for posterior approximation, and $\Theta^{\left(m\right)}$ denotes the $m$-th sample in the stationary Markov chain. Therefore, the posterior distributions of OD vectors are approximated by the set of samples $\left\{\mathcal{Y}^{\left(m\right)}\right\}_{m=1}^M$ during the sampling process, where $\mathcal{Y}^{\left(m\right)}$ denotes the $m$-th sample in the Markov chain.

\section{Experiments}
\label{sec:exp}
Here we conduct numerical experiments using real-world data to evaluate the performance of our proposed model.

\subsection{Data and Experiment Settings}

\begin{figure}[!t]
\centering
\includegraphics[width = 0.95\textwidth]{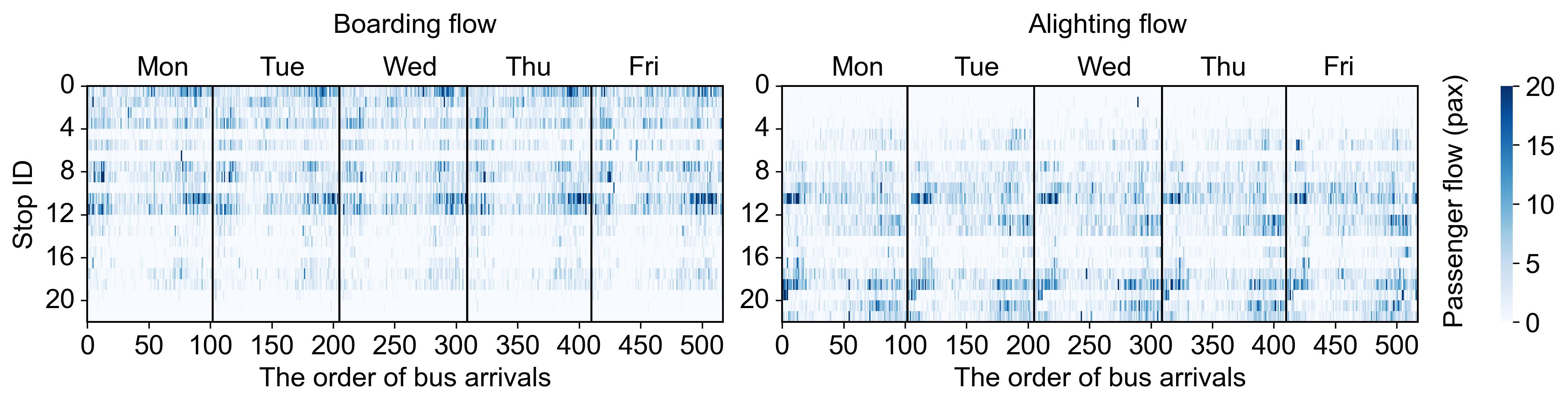}
\caption{One-week boarding and alighting passenger flows of buses at stops. There are two panels: the left shows the boarding counts and the right presents the alighting counts. The x-axis represents different bus IDs and the y-axis represents the stop IDs. The color indicates the volume of passenger flows.}
\label{fig:ba_flow}
\end{figure}

\begin{figure}[!t]
\centering
\includegraphics[width = 0.95\textwidth]{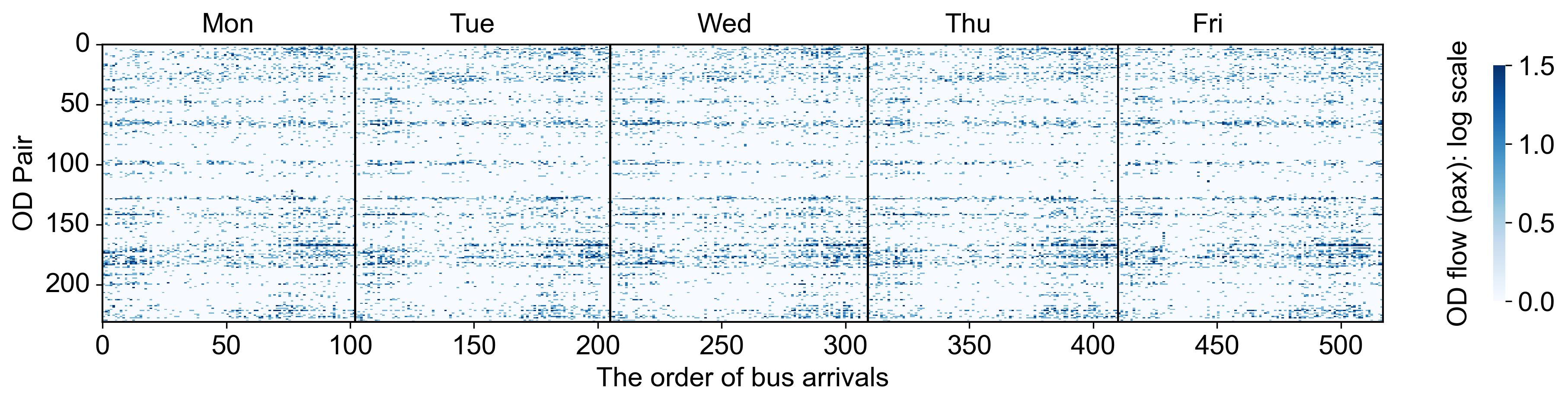}
\caption{One-week OD vectors/flows of all buses. The x-axis represents different bus IDs and the y-axis represents the OD pair IDs. The color indicates the volume of flows.}
\label{fig:od_pair}
\end{figure}

To evaluate our approach, we use high-quality AFC data from three distinct bus routes in a city---a short route with 22 stops, a medium route with 40 stops, and a long route with 72 stops. These bus routes are in operation daily between 6:00 AM and 11:00 PM. The short route operates 103 bus runs daily with a peak frequency of 8.5 buses/hour and an off-peak frequency of 5.5 buses/hour; the medium route operates 85 bus runs daily with a peak frequency of 7.0 buses/hour and an off-peak frequency of 4.5 buses/hour; the long route operates 68 bus runs daily with a peak frequency of 6.0 buses/hour and an off-peak frequency of 4.0 buses/hour. For each bus journey, the AFC data allows us to reconstruct the true OD matrices, serving as the ground truth. We obtain boarding/alighting counts based on the true OD matrices and then apply the proposed model to infer/estimate OD matrices based on the counts. This enables us to directly evaluate the performance of our proposed model alongside other baseline methods by comparing the estimated OD matrices with the ground truth. Fig.~\ref{fig:ba_flow} visualizes the boarding/alighting counts at all stops over one week (from Monday to Friday). Notably, the data reveals significant fluctuations in passenger counts within a day, delineating distinct peak and off-peak hours. Furthermore, a clear daily periodicity is evident over the course of the week. Fig.~\ref{fig:od_pair} presents the one-week OD vectors/flows. These passenger counts exhibit similar temporal patterns as observed in the boarding/alighting counts. Both figures demonstrate the time-varying structure of passenger demand and OD matrices, highlighting the importance of considering temporal dynamics in the estimation of OD matrices.

\subsection{Model Comparison}
In this paper, we compare the performance of our model with the method proposed by \cite{hazelton2010statistical}, which we consider to be a state-of-the-art approach in transit OD matrix estimation at the bus-journey level. Hazelton's method assumes that OD passenger flow follows a Poisson distribution with a travel rate parameter, which is then parameterized by three parameters with Gamma priors. Consequently, we refer to Hazelton's approach as the Poisson-Gamma model in our comparisons.

In addition, we also compare the performance of our model with the widely used IPF method \citep{moshe1037alternative}, which is commonly employed in practice for transit OD matrix estimation. Here, we briefly describe the IPF algorithm as follows:
\begin{itemize}
    \item Reference/seed matrix: We divide one day into four periods, i.e., morning peak, midday off-peak, afternoon peak, and evening off-peak hours. For each period, we randomly select three true OD matrices and calculate the average OD matrix as the reference matrix. For a more accurate comparison, we could refine the seed matrix to address these structural zeros before running the IPF method. Here, we add a small value of 0.01 to the seed matrix before applying the IPF method. This adjustment helps prevent the occurrence of structural zeroes, which can distort the OD matrix estimation. By making this modification, we ensure a more accurate and fair comparison between the IPF method and our proposed model. This small adjustment is a common practice to improve the robustness of the IPF solution, especially in cases where the seed matrix may contain many zero values. The seed matrices are initial estimates of OD matrices. 
    \item Scaling rows and columns: The IPF algorithm iteratively scales the rows and columns of the reference matrix to match the observed boarding and alighting counts for each bus journey. Let $\boldsymbol{Y}$ be the initial reference matrix. In each iteration, the rows of $\boldsymbol{Y}$ are scaled so that their sums match the elements of $\boldsymbol{u}$, and then the columns of $\boldsymbol{Y}$ are scaled to match the elements of $\boldsymbol{v}$. This row and column scaling can be represented as follows:
        \begin{align}
            {y}_{i,j}^{\left(\text{new}\right)} &= {y}_{i,j}\frac{u_i}{\sum_jy_{i,j}}\quad\quad\left(\text{row scaling}\right),\\{y}_{i,j}^{\left(\text{new}\right)} &= {y}_{i,j}\frac{v_j}{\sum_iy_{i,j}}\quad\quad\left(\text{column scaling}\right).
        \end{align}
    \item Convergence: The process iterates until convergence, i.e., when the change between two iterations is below a predetermined threshold.
\end{itemize}

We apply this IPF method to estimate all OD matrices for different periods and then compare the performance to our proposed model. In this work, we use the root mean squared error (RMSE) to evaluate the performance of point estimation based on the mean. We use the continuous rank probability score (CRPS) to evaluate the probabilistic estimation.

\begin{itemize}
    \item  RMSE is defined as:
    \begin{equation}
    \begin{aligned}
         \text{RMSE}&=\sqrt{\frac{1}{n}\sum_{i=1}^{n}(y_i-\hat{y}_{i})^2}, 
    \end{aligned}
    \end{equation}
    where $y_i,\hat{y}_{i},i=1,\ldots,n$ are the true values and estimates, respectively.
    \item The continuous rank probability score is often used as a quantitative measure of probabilistic estimation; it is defined as the quadratic measure of the discrepancy between the estimated cumulative distribution function (CDF, denoted by $F_{X}$) and $\mathbb{I}(x\geq y)$, the empirical CDF of the observation $y$:
    \begin{equation}
        \text{CRPS}=\int_{-\infty}^{\infty}\left(F_{X}(x)-\mathbb{I}(x\geq y)\right)^2dx,
    \end{equation}
    where $\mathbb{I}(\cdot)$ is the indicator function, which is defined as follows: If the condition inside the parentheses is true, then $\mathbb{I}(\cdot)$ equals 1. In many applications, the distribution of estimates is not available in analytic form, but only through simulated samples $X_1,\ldots,X_n$. In this case, the CRPS can be calculated by
    \begin{equation}
         \text{CRPS}=\frac{1}{n}\sum_{i=1}^n\left|X_i-y\right|-\frac{1}{2n^2}\sum_{i=1}^n\sum_{j=1}^n\left|X_i-X_j\right|.
    \end{equation}
    We use the average CRPS of all observations as one metric. Essentially, the CRPS calculates the mean squared difference between the estimated probabilities and the observed outcome, integrated over all possible threshold values. Lower CRPS values indicate better estimation.
\end{itemize}

\subsection{Estimation Results}
\begin{figure}[!t]
\centering
\includegraphics[width = 0.98\textwidth]{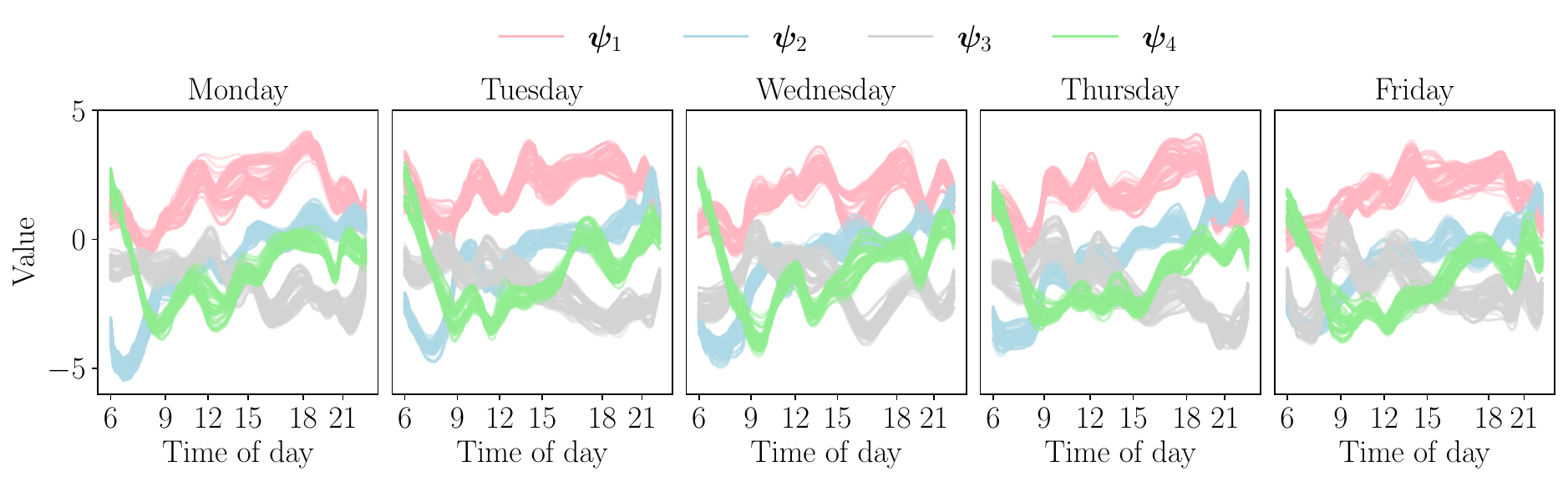}
\caption{Posterior samples of $\boldsymbol{\Psi}$ with rank $D=4$. We show 100 samples of $\boldsymbol{\Psi}$. Samples of different columns are plotted in different colors.}
\label{fig:Psi_samples}
\end{figure}

We implement the developed MCMC algorithm and run a total of 100,000 iterations to sample the model parameters. The code was developed and implemented using Python 3.11 on a laptop with an 11th Gen Intel(R) Core(TM) i9-11950H @ 2.60GHz processor and 32 GB DDR3 RAM. The run times for 10,000 iterations were 3,063 seconds for the short route, 3,447 seconds for the medium route, and 6,119 seconds for the long route. These run times demonstrate that the proposed method is computationally feasible and practical for real-world applications, even for longer routes with more stops. We use the first 95,000 iterations as ``burn-in'' and the last $M_2=5,000$ iterations to approximate the posterior distributions. As an example, Figure~\ref{fig:Psi_samples} shows the 100 randomly selected samples of $\boldsymbol{\Psi}$ with rank $D=4$. Although we use a simple squared-exponential kernel function in the prior, the posterior samples of $\boldsymbol{\Psi}$ still show a clear daily periodic pattern, which confirms the consistent time-dependent characteristics of travel demand. While the temporal factors may appear to have correlations in their temporal tendencies, they are not strictly correlated. In addition, the temporal factors are not required to be independent. We impose a Gaussian process prior on the temporal factors, which naturally allows for correlations between them. In fact, correlated temporal factors are commonly observed in matrix factorization approaches to spatiotemporal data analysis \citep[e.g.,][]{yu2016temporal,chen2021bayesian}. It is common for real-world spatiotemporal data to exhibit a certain degree of correlation. We selected the rank 
$D$ based on the log-likelihood criterion, ensuring that the chosen rank optimally balances the model's ability to fit the data while reducing computational complexity. 

To demonstrate the importance of integrating temporal dynamics in OD matrix estimation, we compare the performance of the temporal Bayesian model with a non-temporal variant. The non-temporal approach assumes static parameters and is derived from our model with rank $D=1$ and $\boldsymbol{\Psi}=\boldsymbol{1}_{N\times 1}$. This ensures that the $N$ journeys share the same alighting probabilities. 
We evaluate the log-likelihood of true OD matrices given the estimated multinomial parameters. A larger log-likelihood value signifies a better model. Table~\ref{tab: od_performance} presents the log-likelihood of different models for the estimation of OD matrices. For the temporal Bayesian model, we implement four variants with different ranks (1, 2, 4 and 6). First, we compare the static model with the temporal model with $D=1$. The key difference between these two models is the assumption of $\boldsymbol{\Psi}$---the static model defines $\boldsymbol{\Psi}$ as a column vector of ones, while the temporal model treats $\boldsymbol{\Psi}$ as a random vector generated from a Gaussian process. From the results, we can see that having a temporal component (even with $D=1$) can greatly enhance the quality of the model, confirming the importance of designing a model with time-varying parameters for OD estimation. In addtion, we can see that the log-likelihood evalutions increase with the rank for all the three bus routes. However, while more factors can enhance model performance, the improvement becomes rather marginal when $D\ge 4$. We use the results from models with rank $D=4$ in the following analysis.

\begin{table}[!t]\centering
\caption{Log-likelihood of different models for OD matrices estimation.}\label{tab: od_performance}
\scriptsize
\begin{tabularx}{\textwidth}{@{}c|c|Y|Y|Y|Y|Y@{}} 
\toprule
\multicolumn{2}{c|}{\multirow{2}{*}{}} & \multirow{2}{*}{\makecell{Static model}} & \multicolumn{4}{c}{Temporal Bayesian model}\\
\cmidrule{4-7}
\multicolumn{1}{c}{}&{}&{}&$D=1$&$D=2$&$D=4$&$D=6$\\
\midrule
\multirow{2}{*}{Short route} & Mean & -35328.77 & -34829.07 & -33728.49 & -33064.69 & -32874.45 \\
{} & Standard deviation  & 98.63  & 86.65 & 79.14 & 86.75 & 85.54\\
\midrule
\multirow{2}{*}{Medium route} & Mean & -63599.00 & -62915.15  & -62141.88 & -61652.34 & -61539.26\\
{} & Standard deviation & 165.46  & 150.24 & 134.89 & 156.80 & 123.00 \\
\midrule
\multirow{2}{*}{Long route} & Mean & -47449.85 & -47078.95 & -46179.38 & -45722.42 & -45722.37\\
{} & Standard deviation & 158.43 & 139.10 & 134.59 & 134.88 & 124.12\\
\bottomrule
\end{tabularx}
\end{table}

\begin{figure}[!t]
\centering
\includegraphics[width = 0.95\textwidth]{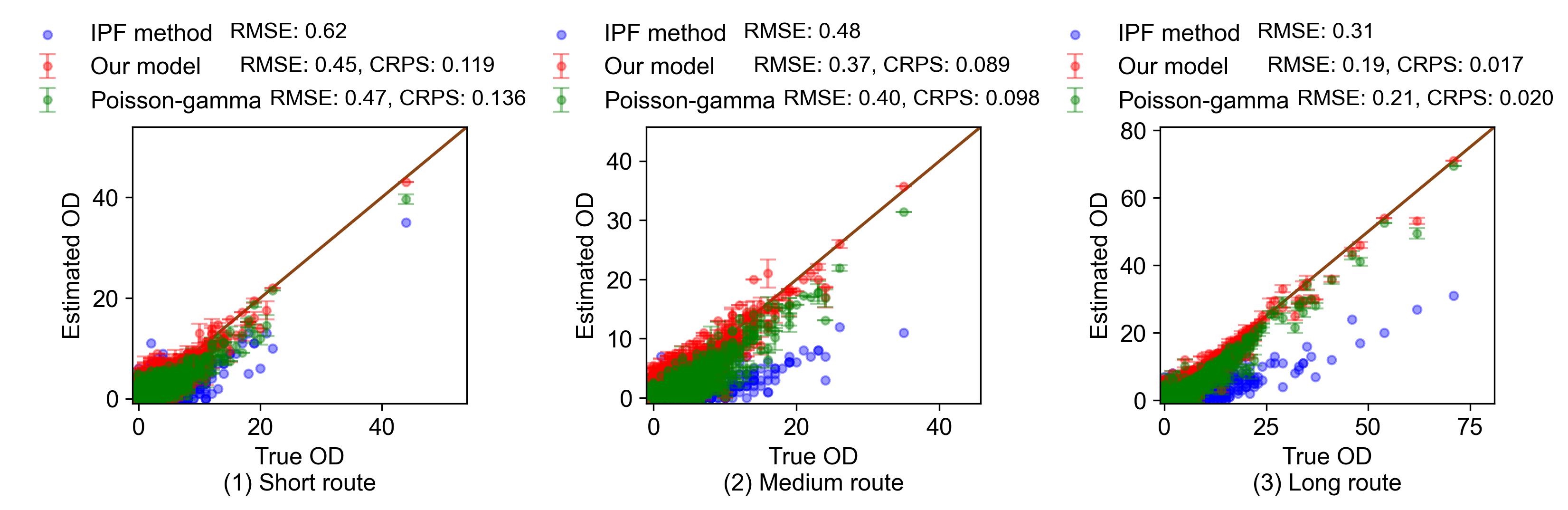}
\caption{True and estimated OD flow of our proposed and baseline models for different routes. These three scatter plots illustrate the actual versus estimated OD flows for the short, medium, and long bus routes. The red scatters represent the estimates from our proposed model, while the blue and green scatters correspond to the IPF method and the Poisson-Gamma model, respectively. Each red and green scatter is accompanied by an error bar that indicates the standard deviation of the estimation, providing a visual representation of the uncertainty associated with each method. At the top of each plot, we present the RMSE and CRPS for each method, offering a quantitative measure of their performance in estimating the OD flows.}
\label{fig:performance}
\end{figure}

\begin{figure}[!t]
\centering
\includegraphics[width = 0.98\textwidth]{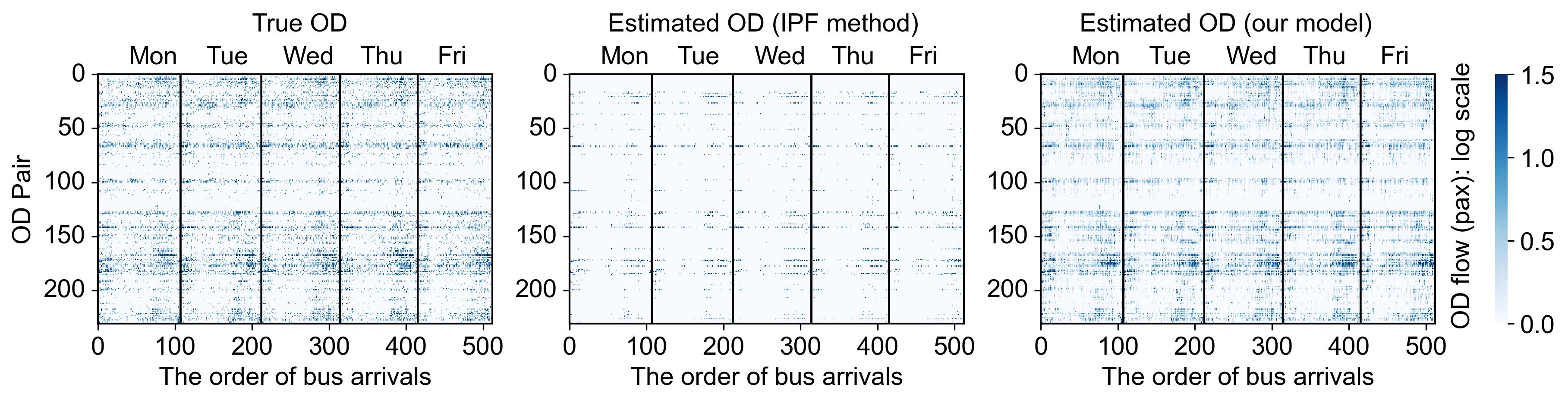}
\caption{True and estimated OD vectors of all buses on the short bus route. There are three panels displaying the OD vectors/flows of all buses: the first is the true OD flows, the second is estimated with the IPF method, and the last is derived from our proposed model. The x-axis represents different bus IDs and the y-axis represents the OD pair IDs. The color represents the volume of passenger flows.}
\label{fig:OD_all_flow}
\end{figure}

Fig.~\ref{fig:performance} presents true and estimated OD flows at the journey level derived from the baseline methods and our model for the three routes. Because IPF is a deterministic method, for the Bayesian method, we use the posterior mean as the estimated demand for model evaluation.
The diagonal line of each plot presents the reference line of perfect estimation, where the estimated flows would align exactly with the true flows. The method with the dots closer to the reference line gives more accurate estimation. We can observe that the dots obtained from our model are closer to the reference line for all bus routes, indicating that our model outperforms the IPF method and the Poisson-Gamma model. Each red and green scatter includes an error bar, representing the standard deviation of the estimation, which provides a visual indication of the uncertainty inherent in our proposed method and Poisson-Gamma model. Overall, both our Bayesian model and the Poisson-Gamma model demonstrate robust estimation capabilities. Moreover, we use RMSE and CRPS to compare the performance of our Bayesian model and the baseline methods. We observe that both the proposed model and the Poisson-Gamma model achieve significantly lower RMSE values compared to those obtained from the IPF method, demonstrating their superior performance. Furthermore, when comparing our model to the Poisson-Gamma model, our proposed method shows a slight but consistent improvement in estimation accuracy. Fig.~\ref{fig:OD_all_flow} shows the estimated OD vectors of all buses with the IPF method and our model on the short route. The results of the medium and long routes are shown in Appendix \ref{app:all od}. Upon inspection, the OD vectors estimated by our model exhibit a closer resemblance to the true OD vectors compared to those generated by the IPF method. This observation underscores the superior performance of our model over the IPF method in accurately capturing and representing the OD flows.

\begin{figure}[!t]
\centering
\includegraphics[width = 0.95\textwidth]{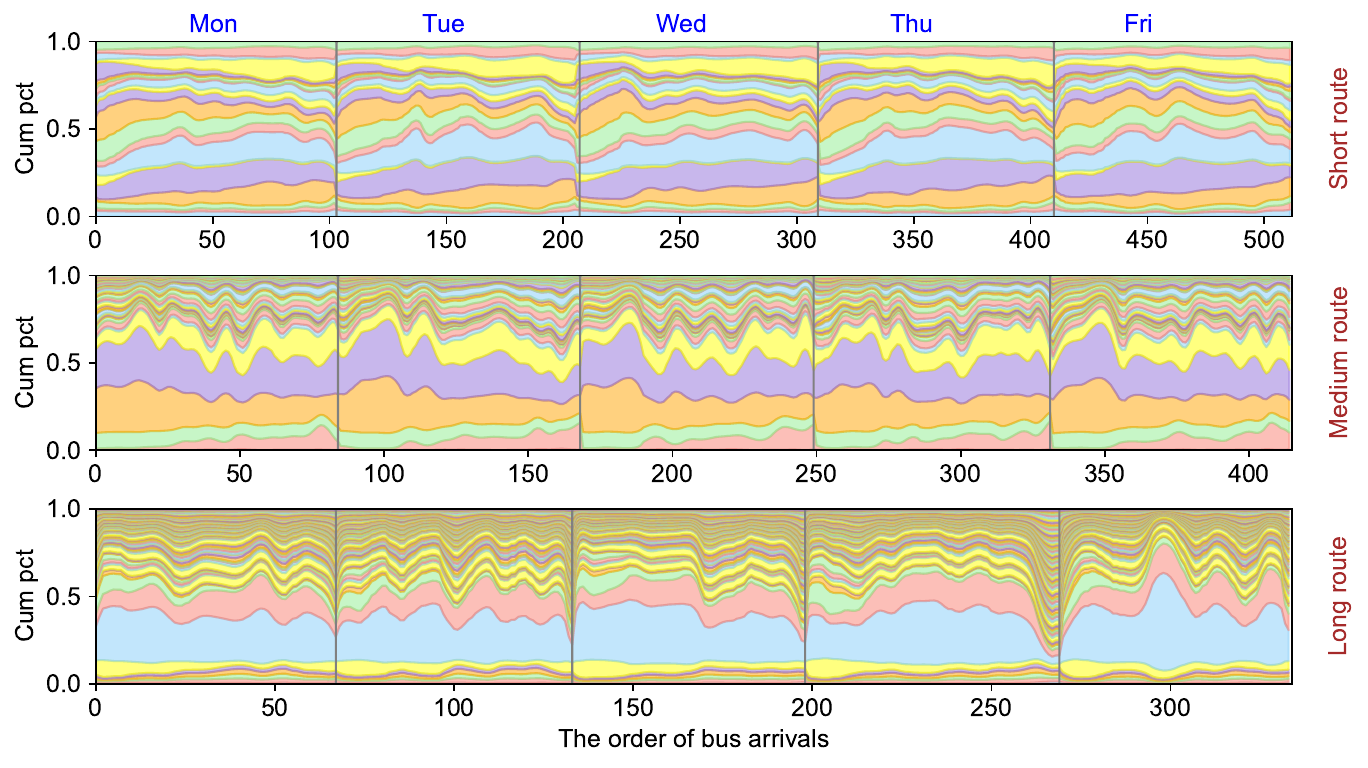}
\caption{Temporal patterns of $\boldsymbol{\lambda}_1$ for different bus routes. There are three panels for different bus routes. Each panel presents the alighting probabilities of downstream stops (from 2 to $S$) for a passenger boarding at the initial stop. The cumulative alighting probabilities of the stops are shown in order (i.e., stop 2 to stop $S$) from the bottom to the top.}
\label{fig:temporal}
\end{figure}

In Fig.~\ref{fig:temporal}, we visualize how the posterior mean of $\{\boldsymbol{\lambda}_1^n\}$, i.e., the vector of alighting probabilities for the first stop, varies with the sequence of journeys. As can be seen, the parameters show clear time-varying characteristics with substantial differences from morning to evening. Moreover, we can observe that while there are slight day-to-day variations in the parameters, the temporal patterns of the parameters exhibit clear similarity/periodicity across days, which is consistent with the estimate of $\boldsymbol{\Psi}$ (see Fig.~\ref{fig:Psi_samples}).

\begin{figure}[!ht]
\centering
\includegraphics[width = 0.95\textwidth]{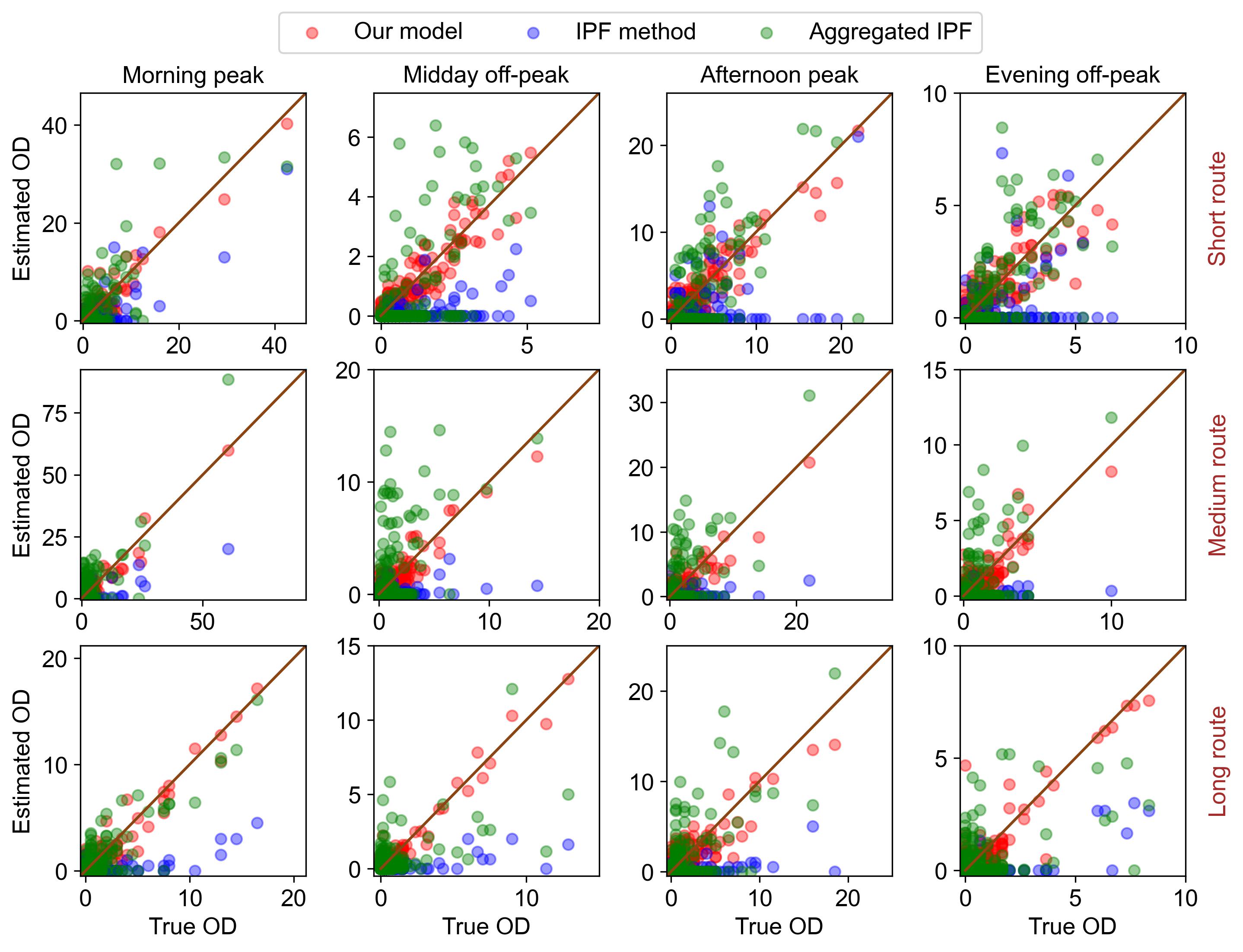}
\caption{True and estimated average hourly OD flow of IPF and our model for different routes. The panels are shown in three rows (represent the short, medium, and long routes, respectively) and four columns (represent the period of morning peak, midday off-peak, afternoon peak, and evening off-peak, respectively). Each panel shows the actual and estimated average hourly OD flow of a route during a specific period. Red scatters are estimated from our proposed model, blue scatters are from the IPF method, and green scatters are from the aggregated IPF method.}
\label{fig:OD_hour_flow}
\end{figure}

In addition to journey-level analysis, transit agencies often use aggregated OD matrices during a certain time period as a proxy for travel demand, serving as input for downstream operational tasks such as timetabling and fleet allocation. To get temporally aggregated OD matrices, we can simply aggregate those OD matrices derived from each bus journey over a defined time window. In practice, the period/window of interest by operators typically includes morning peak (commuting for work/school trips), midday off-peak, afternoon peak (commuting for home trips), and evening off-peak. Specifically, we define the periods/windows as morning peak (7:00--9:00 AM), midday off-peak (9:00 AM--5:00 PM), afternoon peak (5:00--7:00 PM), and evening off-peak (7:00--11:00 PM). We aggregate the estimated OD matrices of all buses into OD matrices of the four periods and further calculate the average hourly OD matrices of the periods to validate the performance of our model. For aggregated OD matrices, the journey-based IPF model is expected to perform poorly due to overfitting. A more appropriate benchmark is to fit a single IPF model for each period using aggregated marginal counts. We refer to this approach as ``aggregated IPF''. Fig.~\ref{fig:OD_hour_flow} provides aggregated hourly OD matrices for different periods for all bus routes. These scatter plots present the hourly variability in performance and reveal temporal patterns in the estimation accuracy of our model. We can see that our model outperforms aggregated IPF and achieves accurate estimations of different periods for all bus routes.

\begin{figure}[!t]
\centering
\includegraphics[width = 0.95\textwidth]{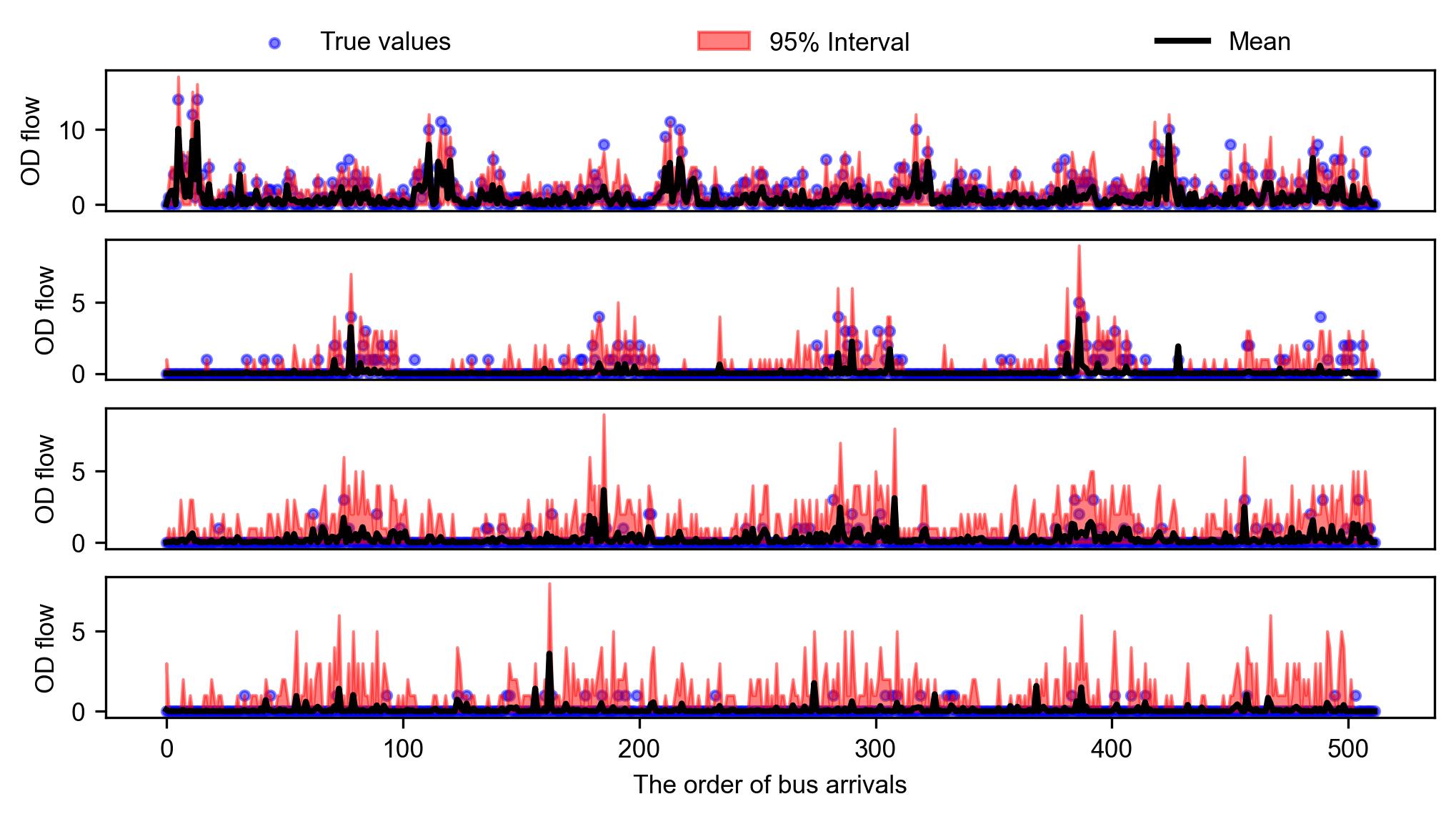}
\caption{Estimation with the uncertainty of some OD pairs on the short bus route. There are four panels and each panel is for a specific OD pair. The x-axis represents the bus IDs and the y-axis represents the OD flow. The blue dots represent the true OD flows of buses, the black line represents the estimated mean of the OD flows, and the red shade areas represent the estimated 95\% confidence intervals.}
\label{fig:est_uncer}
\end{figure}

A unique advantage of Bayesian inference is that we can get the posterior distributions associated with each entry in the journey-level OD matrix. Fig.~\ref{fig:est_uncer} presents estimations with uncertainties (95\% credible interval) of all buses on the short route. The results of the medium and long routes are shown in Appendix \ref{app:uncertainty}. As can be seen, the presented entries indeed vary substantially over the sequence of journeys. This further supports our choice of using the alighting probability instead of demand intensity to build the time-varying model. We can see that the performance of the estimations varies across different buses and OD pairs. In most cases, we observe good uncertainty quantification where the true values align closely with the mean estimates and fall within the 95\% credible intervals. Overall, the model demonstrates robust estimation results with high-quality uncertainty quantification.

\section{Conclusion and Discussion}
\label{sec:con}

In this paper, we propose a novel temporal Bayesian model for inferring transit OD matrices at the individual bus journey level based on boarding/alighting counts at each stop. Given a boarding stop, we model the number of alighting passengers at subsequent bus stops with a multinomial distribution parameterized by a vector of alighting probabilities, and we assume that these probabilities vary smoothly over time. Given the scale of the problem, we design an efficient and effective parameterization using a matrix factorization model with a mapping factor matrix and a temporal factor matrix. In particular, we impose Gaussian process priors on the temporal factor matrix, thus ensuring temporal smoothness in the estimated alighting probabilities. For model inference, we develop an efficient two-stage algorithm based on the MCMC method. Our approach can effectively capture the dynamic nature of OD matrices and bypass the exhaustive enumeration of feasible OD matrices which aligns with observed boarding and alighting counts. We evaluate the proposed model using real-world data, and the results confirm its effectiveness in terms of accurate OD matrix estimation and robust uncertainty quantification.

Our proposed method differs from the work by \cite{hazelton2010statistical} in three key aspects. Firstly, \citet{hazelton2010statistical} used the Poisson distribution to model the passenger counts in their work. However, the Poisson distribution is not well-suited for modeling the passenger counts of bus trips due to the complex characteristics of bus operations, such as bus bunching. In contrast, our work considers each row in the OD matrix as a realization from a multinomial distribution and treats the alighting probabilities for each bus stop as model parameters. This approach is more reasonable and better suited to the nature of bus passenger count data. Secondly, we extend \citet{hazelton2010statistical}  from inferring static demand to time-varying demand. Our model explicitly captures the temporal variations in OD flows by employing a Bayesian framework with time-varying parameters. This allows for a more accurate and realistic representation of passenger demand over time. Thirdly, to effectively model the temporal dynamics with reduced computational complexity, we introduced softmax parameterization of the alighting probabilities and a matrix factorization reparameterization, where the temporal factor matrix follows a Gaussian process prior. This innovation ensures smoothly time-varying alighting probabilities and significantly enhances the model’s ability to handle limited data scenarios, providing a robust framework for capturing time-varying passenger demand.

Our approach has potential implications for both practice and research. First, the proposed temporal Bayesian model can produce posterior distributions for transit OD matrices. It is important to note that distributions are more valuable than point estimates. This is because the associated uncertainty in travel demand distributions could benefit many downstream operational tasks, such as network design and service scheduling, where it is important to make decisions for a range of possible scenarios. Second, we find that the inferred model parameters are highly interpretable. The learned patterns could help agencies better understand how travel demand varies spatially and temporally and further improve the design of transit networks. Third, the estimated alighting probabilities from our model can be leveraged to estimate OD matrices for bus journeys where boarding and alighting counts are not available. Here, we provide a possible specific method. Let $\delta_j^t$ represent the passenger arrival rate at stop $j$ during time period $t$. For a bus $i$ arriving at stop $j$ during period $t$, the boarding count $u_{ij}$ is modeled as $u_{ij}\sim \text{Poisson}\left(h_{ij}\delta_j^t\right)$, where $h_{ij}$ is the headway between bus $i$ and its predecessor at stop $j$. To account for variations across $T$ periods, we define the arrival rate vector for stop $j$ as $\boldsymbol{\delta}_j=\left[\delta_j^1,\ldots,\delta_j^T\right]^\top\in \mathbb{R}^{T}$ and parameterize it as $\boldsymbol{\delta}_j=\text{exp}\left(\boldsymbol{\vartheta}_j\right)$. A Gaussian process prior is imposed on $\boldsymbol{\vartheta}_j$ to ensure smooth temporal variation, with covariance matrix defined by a kernel function. This allows us to estimate the arrival rate vector $\boldsymbol{\delta}_j$ for each stop $j$ and infer the boarding counts $u_{ij}$ using observed headways. Once the boarding counts are estimated, they are combined with the alighting probabilities derived from our model to estimate the OD matrix $\boldsymbol{Y}_i$ for the bus journey. Specifically, each row of the OD matrix is computed as the product of the boarding count and the corresponding alighting probability vector. This yields $\boldsymbol{Y}_{ij}=u_{ij}\boldsymbol{\Lambda}_{ij}$, where $\boldsymbol{\Lambda}_{ij}$ is the $j$-th row of the alighting probability matrix estimated by our model. Additionally, the column sums of $\boldsymbol{Y}_i$ provide the estimated alighting counts for the bus.


It shoud be noted that our model assumes that the APC data is error-free or meets specific constraints, specifically that all boarding and alighting counts are non-negative and that passenger occupancy on each segment is non-negative as well. These constraints are crucial because the model relies on accurate boarding and alighting counts to sample the OD matrix effectively. If the APC data does not adhere to these constraints---due to errors, gaps, or inconsistencies---the model will be unable to accurately sample and estimate the OD matrix, leading to potential failures in the estimation process. To address this issue, it is important to correct any errors or inconsistencies in the APC data before applying our model. This can involve ensuring that the data meets the necessary constraints, such as non-negative counts and valid occupancy levels. There are established methods for correcting APC data in the literature \citep[e.g.,][]{furth2005making,barabino2014offline,ji2015transit,roncoli2023estimating}, which focus on identifying and rectifying data inaccuracies. Implementing such data correction techniques is essential to ensure that the APC data is suitable for use in our model, thereby enabling reliable and accurate OD matrix estimation.
While our approach is designed to effectively infer OD matrices for bus trips under typical operating conditions, it is important to acknowledge certain limitations in its application. Specifically, the method relies on data collected over several consecutive weeks, which helps to capture the regular and smoothly time-varying patterns in passenger behavior. However, this approach does not account for irregular events such as special occasions, seasonal changes, or significant disruptions like major traffic accidents, all of which can markedly influence bus operations and passenger behavior. In scenarios where these factors are significant, additional methods or adjustments may be required to accurately reflect the true dynamics of passenger behavior. Future work could explore integrating or developing complementary approaches to address this specific challenge. Moreover, we could extend the proposed model to incorporate more prior information from additional data sources such as AFC data, which might improve the accuracy of OD matrices inference. In addition, we recognize that applying our proposed approach to an entire bus network, which comprises numerous routes, could present significant computational challenges. These challenges arise from the large volume of data and the necessity to process multiple routes simultaneously. The high computational demand could make it difficult to implement the method efficiently on a large scale. To mitigate this issue, one potential solution is to employ parallel computing, which would allow each bus route to be processed independently. This approach could significantly reduce computational time, making the method more practical for large-scale applications across entire transit networks. In future work, we plan to explore and implement these optimizations to enhance the scalability and feasibility of our approach in real-world settings. Lastly, while our current model employs a Markov proposal for sampling the OD matrix, we recognize that this effective approach may not be the most efficient option available. To enhance the efficiency of the sampling process, more advanced global sampling techniques could be considered. One such approach is the dynamic Markov basis \citep{hazelton2021geometrically}, which offers a more efficient global sampling mechanism. This method could potentially outperform the current Markov proposal by exploring the sample space more thoroughly and reducing the convergence time in the MCMC process. In future work, we plan to explore the integration of dynamic Markov basis sampling into our methodology. This enhancement would not only improve the computational performance of the model but also potentially increase the accuracy and robustness of the OD matrix estimates, particularly in scenarios with more intricate passenger flow patterns.

\section*{Acknowledgements}

This research is supported in part by the Canadian Statistical Sciences Institute (CANSSI) Collaborative Research Teams (CRT) grants and in part by the Natural Sciences and Engineering Research Council (NSERC) of Canada. X. Chen acknowledges the funding
support from the China Scholarship Council (CSC).

\bibliographystyle{elsarticle-harv}
\bibliography{Reference}

\begin{thebibliography}{46}
\providecommand{\natexlab}[1]{#1}
\providecommand{\url}[1]{\texttt{#1}}
\expandafter\ifx\csname urlstyle\endcsname\relax
  \providecommand{\doi}[1]{doi: #1}\else
  \providecommand{\doi}{doi: \begingroup \urlstyle{rm}\Url}\fi

\bibitem[Agrawal et~al.(2017)Agrawal, Granger-Bevan, Newmark, and
  Nixon]{agrawal2017comparing}
Asha~Weinsten Agrawal, Stephen Granger-Bevan, Gregory~L Newmark, and Hilary
  Nixon.
\newblock Comparing data quality and cost from three modes of on-board transit
  surveys.
\newblock \emph{Transport Policy}, 54:\penalty0 70--79, 2017.

\bibitem[Ahern et~al.(2022)Ahern, Paz, and Corry]{ahern2022approximate}
Zeke Ahern, Alexander Paz, and Paul Corry.
\newblock Approximate multi-objective optimization for integrated bus route
  design and service frequency setting.
\newblock \emph{Transportation Research Part B: Methodological}, 155:\penalty0
  1--25, 2022.

\bibitem[Ashok and Ben-Akiva(2002)]{ashok2002estimation}
Kalidas Ashok and Moshe~E Ben-Akiva.
\newblock Estimation and prediction of time-dependent origin-destination flows
  with a stochastic mapping to path flows and link flows.
\newblock \emph{Transportation Science}, 36\penalty0 (2):\penalty0 184--198,
  2002.

\bibitem[Assemi et~al.(2020)Assemi, Alsger, Moghaddam, Hickman, and
  Mesbah]{assemi2020improving}
Behrang Assemi, Azalden Alsger, Mahboobeh Moghaddam, Mark Hickman, and Mahmoud
  Mesbah.
\newblock Improving alighting stop inference accuracy in the trip chaining
  method using neural networks.
\newblock \emph{Public Transport}, 12:\penalty0 89--121, 2020.

\bibitem[Barabino et~al.(2014)Barabino, Di~Francesco, and
  Mozzoni]{barabino2014offline}
Benedetto Barabino, Massimo Di~Francesco, and Sara Mozzoni.
\newblock An offline framework for handling automatic passenger counting raw
  data.
\newblock \emph{IEEE Transactions on Intelligent Transportation Systems},
  15\penalty0 (6):\penalty0 2443--2456, 2014.

\bibitem[Ben-Akiva et~al.(1985)Ben-Akiva, Macke, and Hsu]{moshe1037alternative}
Moshe Ben-Akiva, Peter~P Macke, and Poh~Ser Hsu.
\newblock Alternative methods to estimate route-level trip tables and expand
  on-board surveys.
\newblock \emph{Transportation Research Record}, 1037:\penalty0 1--11, 1985.

\bibitem[Blei and Lafferty(2006)]{blei2006dynamic}
David~M Blei and John~D Lafferty.
\newblock Dynamic topic models.
\newblock In \emph{Proceedings of the 23rd international conference on Machine
  learning}, pages 113--120, 2006.

\bibitem[Blume et~al.(2022)Blume, Corman, and Sansavini]{blume2022bayesian}
Steffen~OP Blume, Francesco Corman, and Giovanni Sansavini.
\newblock Bayesian origin-destination estimation in networked transit systems
  using nodal in-and outflow counts.
\newblock \emph{Transportation Research Part B: Methodological}, 161:\penalty0
  60--94, 2022.

\bibitem[Carvalho(2014)]{carvalho2014bayesian}
Luis Carvalho.
\newblock A bayesian statistical approach for inference on static
  origin--destination matrices in transportation studies.
\newblock \emph{Technometrics}, 56\penalty0 (2):\penalty0 225--237, 2014.

\bibitem[Cascetta et~al.(1993)Cascetta, Inaudi, and
  Marquis]{cascetta1993dynamic}
Ennio Cascetta, Domenico Inaudi, and Gerald Marquis.
\newblock Dynamic estimators of origin-destination matrices using traffic
  counts.
\newblock \emph{Transportation Science}, 27\penalty0 (4):\penalty0 363--373,
  1993.

\bibitem[Chang and Wu(1994)]{chang1994recursive}
Gang-Len Chang and Jifeng Wu.
\newblock Recursive estimation of time-varying origin-destination flows from
  traffic counts in freeway corridors.
\newblock \emph{Transportation Research Part B: Methodological}, 28\penalty0
  (2):\penalty0 141--160, 1994.

\bibitem[Chen and Sun(2021)]{chen2021bayesian}
Xinyu Chen and Lijun Sun.
\newblock Bayesian temporal factorization for multidimensional time series
  prediction.
\newblock \emph{IEEE Transactions on Pattern Analysis and Machine
  Intelligence}, 44\penalty0 (9):\penalty0 4659--4673, 2021.

\bibitem[Cheng et~al.(2021)Cheng, Tr{\'e}panier, and
  Sun]{cheng2021probabilistic}
Zhanhong Cheng, Martin Tr{\'e}panier, and Lijun Sun.
\newblock Probabilistic model for destination inference and travel pattern
  mining from smart card data.
\newblock \emph{Transportation}, 48\penalty0 (4):\penalty0 2035--2053, 2021.

\bibitem[Cho et~al.(2009)Cho, Jou, and Lan]{cho2009time}
Hsun-Jung Cho, Yow-Jen Jou, and Chien-Lun Lan.
\newblock Time dependent origin-destination estimation from traffic count
  without prior information.
\newblock \emph{Networks and Spatial Economics}, 9:\penalty0 145--170, 2009.

\bibitem[Cremer and Keller(1987)]{cremer1987new}
M~Cremer and H~Keller.
\newblock A new class of dynamic methods for the identification of
  origin-destination flows.
\newblock \emph{Transportation Research Part B: Methodological}, 21\penalty0
  (2):\penalty0 117--132, 1987.

\bibitem[Furth et~al.(2005)Furth, Strathman, and Hemily]{furth2005making}
Peter~G Furth, James~G Strathman, and Brendon Hemily.
\newblock Making automatic passenger counts mainstream: Accuracy, balancing
  algorithms, and data structures.
\newblock \emph{Transportation Research Record}, 1927\penalty0 (1):\penalty0
  206--216, 2005.

\bibitem[Gkiotsalitis et~al.(2019)Gkiotsalitis, Wu, and
  Cats]{gkiotsalitis2019cost}
Konstantinos Gkiotsalitis, Zongxiang Wu, and Oded Cats.
\newblock A cost-minimization model for bus fleet allocation featuring the
  tactical generation of short-turning and interlining options.
\newblock \emph{Transportation Research Part C: Emerging Technologies},
  98:\penalty0 14--36, 2019.

\bibitem[Hazelton(2008)]{hazelton2008statistical}
Martin~L Hazelton.
\newblock Statistical inference for time varying origin--destination matrices.
\newblock \emph{Transportation Research Part B: Methodological}, 42\penalty0
  (6):\penalty0 542--552, 2008.

\bibitem[Hazelton(2010)]{hazelton2010statistical}
Martin~L Hazelton.
\newblock Statistical inference for transit system origin-destination matrices.
\newblock \emph{Technometrics}, 52\penalty0 (2):\penalty0 221--230, 2010.

\bibitem[Hazelton and Parry(2016)]{hazelton2016statistical}
Martin~L Hazelton and Katharina Parry.
\newblock Statistical methods for comparison of day-to-day traffic models.
\newblock \emph{Transportation Research Part B: Methodological}, 92:\penalty0
  22--34, 2016.

\bibitem[Hazelton et~al.(2021)Hazelton, Mcveagh, and
  Van~Brunt]{hazelton2021geometrically}
ML~Hazelton, MR~Mcveagh, and B~Van~Brunt.
\newblock Geometrically aware dynamic markov bases for statistical linear
  inverse problems.
\newblock \emph{Biometrika}, 108\penalty0 (3):\penalty0 609--626, 2021.

\bibitem[Hussain et~al.(2021)Hussain, Bhaskar, and Chung]{hussain2021transit}
Etikaf Hussain, Ashish Bhaskar, and Edward Chung.
\newblock Transit od matrix estimation using smartcard data: Recent
  developments and future research challenges.
\newblock \emph{Transportation Research Part C: Emerging Technologies},
  125:\penalty0 103044, 2021.

\bibitem[Ji et~al.(2014)Ji, Mishalani, and McCord]{ji2014estimating}
Yuxiong Ji, Rabi~G Mishalani, and Mark~R McCord.
\newblock Estimating transit route od flow matrices from apc data on multiple
  bus trips using the ipf method with an iteratively improved base: Method and
  empirical evaluation.
\newblock \emph{Journal of Transportation Engineering}, 140\penalty0
  (5):\penalty0 04014008, 2014.

\bibitem[Ji et~al.(2015)Ji, Mishalani, and McCord]{ji2015transit}
Yuxiong Ji, Rabi~G Mishalani, and Mark~R McCord.
\newblock Transit passenger origin--destination flow estimation: Efficiently
  combining onboard survey and large automatic passenger count datasets.
\newblock \emph{Transportation Research Part C: Emerging Technologies},
  58:\penalty0 178--192, 2015.

\bibitem[Lamond and Stewart(1981)]{lamond1981bregman}
B~Lamond and Neil~F Stewart.
\newblock Bregman's balancing method.
\newblock \emph{Transportation Research Part B: Methodological}, 15\penalty0
  (4):\penalty0 239--248, 1981.

\bibitem[Lei et~al.(2022)Lei, Labbe, Wu, and Sun]{lei2022bayesian}
Mengying Lei, Aurelie Labbe, Yuankai Wu, and Lijun Sun.
\newblock Bayesian kernelized matrix factorization for spatiotemporal traffic
  data imputation and kriging.
\newblock \emph{IEEE Transactions on Intelligent Transportation Systems},
  23\penalty0 (10):\penalty0 18962--18974, 2022.

\bibitem[Li(2009)]{li2009markov}
Baibing Li.
\newblock Markov models for bayesian analysis about transit route
  origin--destination matrices.
\newblock \emph{Transportation Research Part B: Methodological}, 43\penalty0
  (3):\penalty0 301--310, 2009.

\bibitem[Li and Cassidy(2007)]{li2007generalized}
Yuwei Li and Michael~J Cassidy.
\newblock A generalized and efficient algorithm for estimating transit route
  {OD}s from passenger counts.
\newblock \emph{Transportation Research Part B: Methodological}, 41\penalty0
  (1):\penalty0 114--125, 2007.

\bibitem[Lopes et~al.(2008)Lopes, Salazar, and Gamerman]{lopes2008spatial}
Hedibert~Freitas Lopes, Esther Salazar, and Dani Gamerman.
\newblock Spatial dynamic factor analysis.
\newblock \emph{Bayesian Analysis}, 3\penalty0 (4):\penalty0 759--792, 2008.

\bibitem[Luttinen and Ilin(2009)]{luttinen2009variational}
Jaakko Luttinen and Alexander Ilin.
\newblock Variational gaussian-process factor analysis for modeling
  spatio-temporal data.
\newblock \emph{Advances in Neural Information Processing Systems}, 22, 2009.

\bibitem[Mart{\'\i}nez et~al.(2014)Mart{\'\i}nez, Mauttone, and
  Urquhart]{martinez2014frequency}
H{\'e}ctor Mart{\'\i}nez, Antonio Mauttone, and Mar{\'\i}a~E Urquhart.
\newblock Frequency optimization in public transportation systems: Formulation
  and metaheuristic approach.
\newblock \emph{European Journal of Operational Research}, 236\penalty0
  (1):\penalty0 27--36, 2014.

\bibitem[McCullagh(1989)]{mccullagh2019generalized}
Peter McCullagh.
\newblock \emph{Generalized linear models}.
\newblock Chapman and Hall/CRC, 2nd edition, 1989.

\bibitem[Mohammed and Oke(2023)]{mohammed2023origin}
Mohammed Mohammed and Jimi Oke.
\newblock Origin-destination inference in public transportation systems: A
  comprehensive review.
\newblock \emph{International Journal of Transportation Science and
  Technology}, 12\penalty0 (1):\penalty0 315--328, 2023.

\bibitem[Murray and Adams(2010)]{murray2010slice}
Iain Murray and Ryan~P Adams.
\newblock Slice sampling covariance hyperparameters of latent gaussian models.
\newblock \emph{Advances in Neural Information Processing Systems}, 23, 2010.

\bibitem[Murray et~al.(2010)Murray, Adams, and MacKay]{murray2010elliptical}
Iain Murray, Ryan Adams, and David MacKay.
\newblock Elliptical slice sampling.
\newblock In \emph{Proceedings of the thirteenth international conference on
  artificial intelligence and statistics}, pages 541--548. JMLR Workshop and
  Conference Proceedings, 2010.

\bibitem[Pitombeira-Neto et~al.(2020)Pitombeira-Neto, Loureiro, and
  Carvalho]{pitombeira2020dynamic}
Anselmo~Ramalho Pitombeira-Neto, Carlos Felipe~Grangeiro Loureiro, and
  Luis~Eduardo Carvalho.
\newblock A dynamic hierarchical bayesian model for the estimation of
  day-to-day origin-destination flows in transportation networks.
\newblock \emph{Networks and Spatial Economics}, 20:\penalty0 499--527, 2020.

\bibitem[Roncoli et~al.(2023)Roncoli, Chandakas, and
  Kaparias]{roncoli2023estimating}
Claudio Roncoli, Ektoras Chandakas, and Ioannis Kaparias.
\newblock Estimating on-board passenger comfort in public transport vehicles
  using incomplete automatic passenger counting data.
\newblock \emph{Transportation Research Part C: Emerging Technologies},
  146:\penalty0 103963, 2023.

\bibitem[Sherali and Park(2001)]{sherali2001estimation}
Hanif~D Sherali and Taehyung Park.
\newblock Estimation of dynamic origin--destination trip tables for a general
  network.
\newblock \emph{Transportation Research Part B: Methodological}, 35\penalty0
  (3):\penalty0 217--235, 2001.

\bibitem[Sun et~al.(2014{\natexlab{a}})Sun, Jin, Lee, Axhausen, and
  Erath]{sun2014demand}
Lijun Sun, Jian~Gang Jin, Der-Horng Lee, Kay~W Axhausen, and Alexander Erath.
\newblock Demand-driven timetable design for metro services.
\newblock \emph{Transportation Research Part C: Emerging Technologies},
  46:\penalty0 284--299, 2014{\natexlab{a}}.

\bibitem[Sun et~al.(2014{\natexlab{b}})Sun, Tirachini, Axhausen, Erath, and
  Lee]{sun2014models}
Lijun Sun, Alejandro Tirachini, Kay~W Axhausen, Alexander Erath, and Der-Horng
  Lee.
\newblock Models of bus boarding and alighting dynamics.
\newblock \emph{Transportation Research Part A: Policy and Practice},
  69:\penalty0 447--460, 2014{\natexlab{b}}.

\bibitem[Tr{\'e}panier et~al.(2007)Tr{\'e}panier, Tranchant, and
  Chapleau]{trepanier2007individual}
Martin Tr{\'e}panier, Nicolas Tranchant, and Robert Chapleau.
\newblock Individual trip destination estimation in a transit smart card
  automated fare collection system.
\newblock \emph{Journal of Intelligent Transportation Systems}, 11\penalty0
  (1):\penalty0 1--14, 2007.

\bibitem[Vardi(1996)]{vardi1996network}
Yehuda Vardi.
\newblock Network tomography: Estimating source-destination traffic intensities
  from link data.
\newblock \emph{Journal of the American Statistical Association}, 91\penalty0
  (433):\penalty0 365--377, 1996.

\bibitem[Xie and Duthie(2015)]{xie2015excess}
Chi Xie and Jennifer Duthie.
\newblock An excess-demand dynamic traffic assignment approach for inferring
  origin-destination trip matrices.
\newblock \emph{Networks and Spatial Economics}, 15:\penalty0 947--979, 2015.

\bibitem[Yu et~al.(2016)Yu, Rao, and Dhillon]{yu2016temporal}
Hsiang-Fu Yu, Nikhil Rao, and Inderjit~S Dhillon.
\newblock Temporal regularized matrix factorization for high-dimensional time
  series prediction.
\newblock \emph{Advances in Neural Information Processing Systems}, 29, 2016.

\bibitem[Zhao et~al.(2007)Zhao, Rahbee, and Wilson]{zhao2007estimating}
Jinhua Zhao, Adam Rahbee, and Nigel~HM Wilson.
\newblock Estimating a rail passenger trip origin-destination matrix using
  automatic data collection systems.
\newblock \emph{Computer-Aided Civil and Infrastructure Engineering},
  22\penalty0 (5):\penalty0 376--387, 2007.

\bibitem[Zhou and Mahmassani(2007)]{zhou2007structural}
Xuesong Zhou and Hani~S Mahmassani.
\newblock A structural state space model for real-time traffic
  origin--destination demand estimation and prediction in a day-to-day learning
  framework.
\newblock \emph{Transportation Research Part B: Methodological}, 41\penalty0
  (8):\penalty0 823--840, 2007.

\end{thebibliography}
\pagebreak
\appendix
\counterwithin{figure}{section}
\section{True and estimated OD vectors of all buses.}\label{app:all od}
\begin{figure}[!ht]
\centering
\includegraphics[width = 0.8\textwidth]{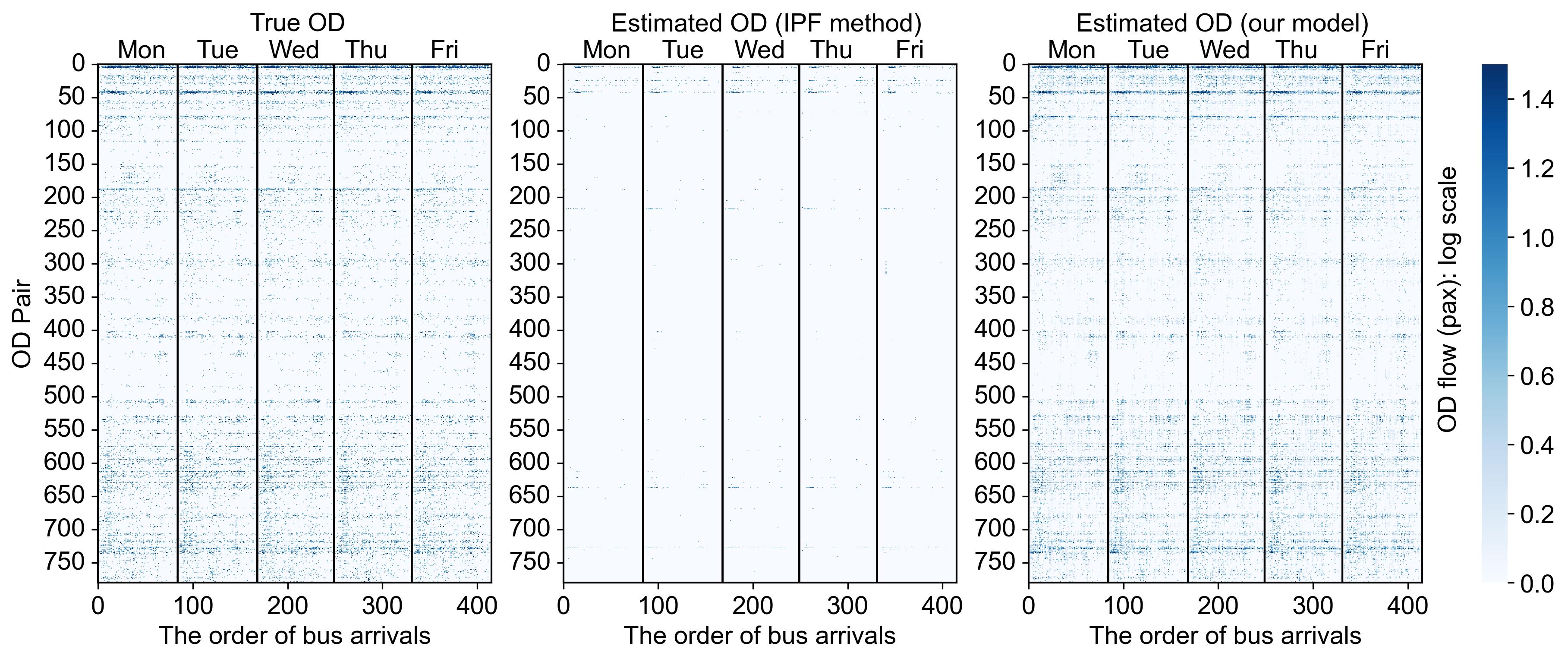}
\caption{True and estimated OD vectors of all buses on the medium bus route.}
\label{fig:medium_OD_all_est}
\end{figure}
\begin{figure}[!ht]
\centering
\includegraphics[width = 0.8\textwidth]{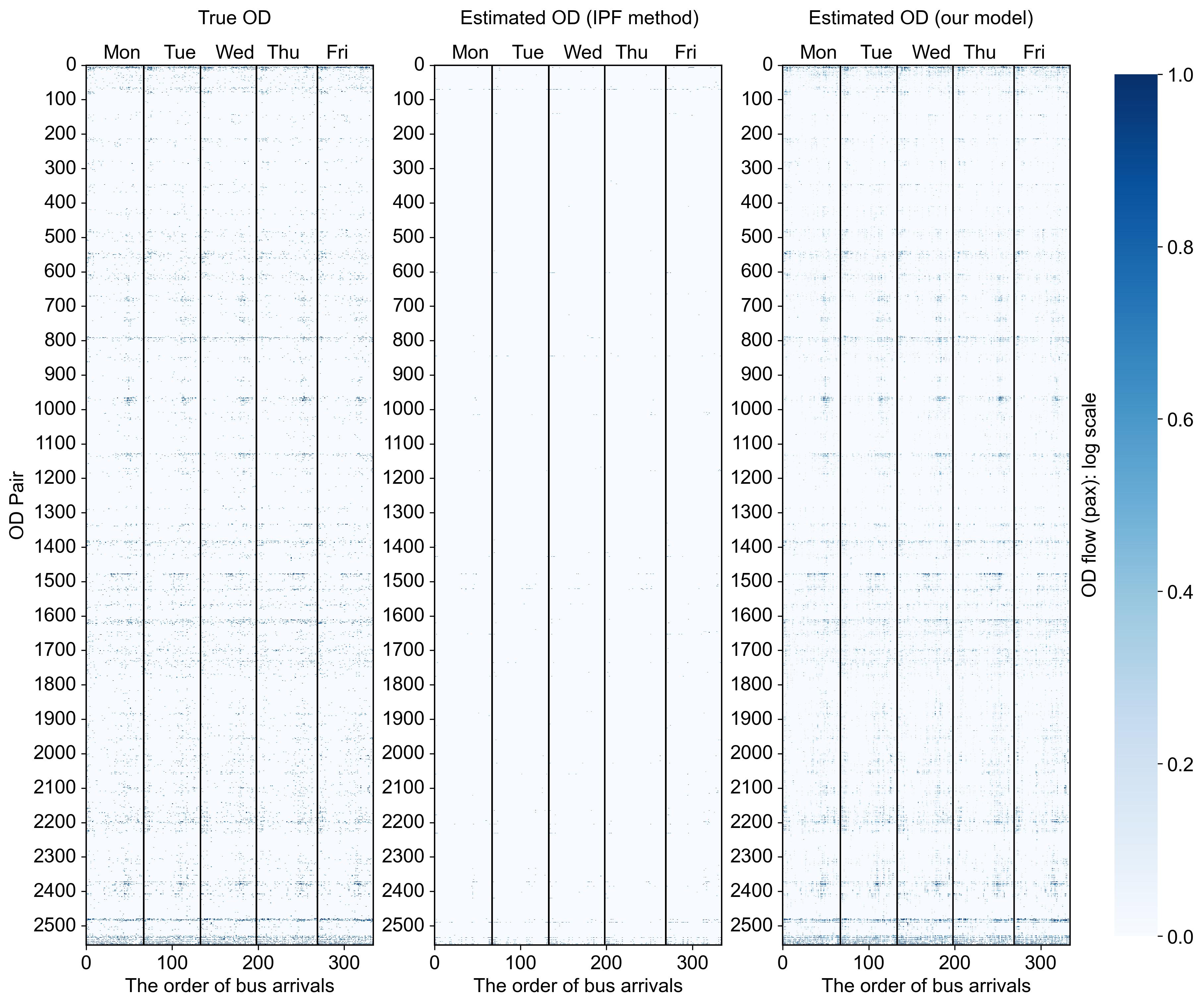}
\caption{True and estimated OD vectors of all buses on the long bus route.}
\label{fig:long_OD_all_est}
\end{figure}

\clearpage
\section{Estimation with uncertainty of some OD pairs.}\label{app:uncertainty}
\begin{figure}[!ht]
\centering
\includegraphics[width = 0.9\textwidth]{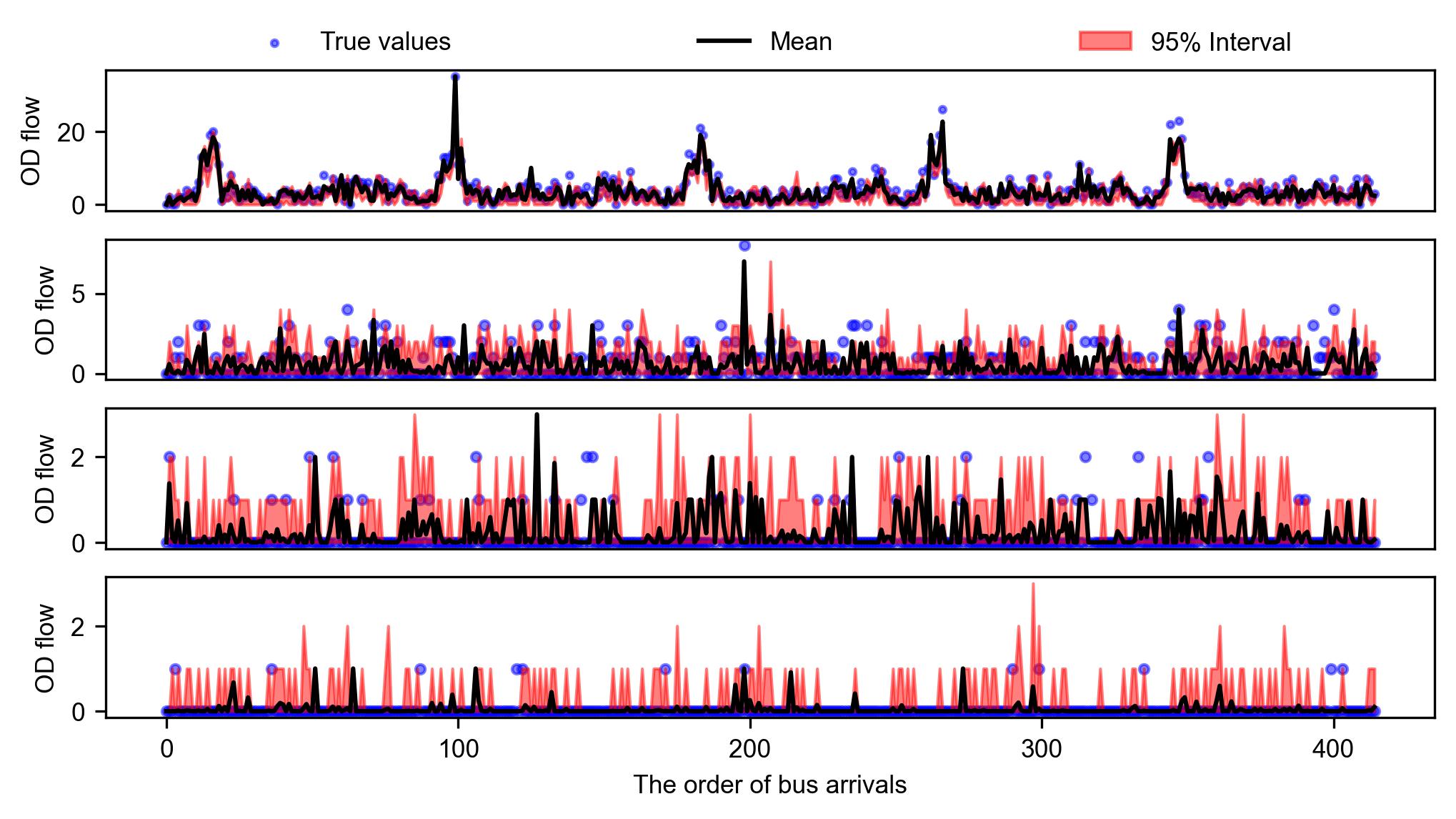}
\caption{Estimation with the uncertainty of some OD pairs on the medium bus route.}
\label{fig:est_uncer_medium}
\end{figure}
\begin{figure}[!ht]
\centering
\includegraphics[width = 0.9\textwidth]{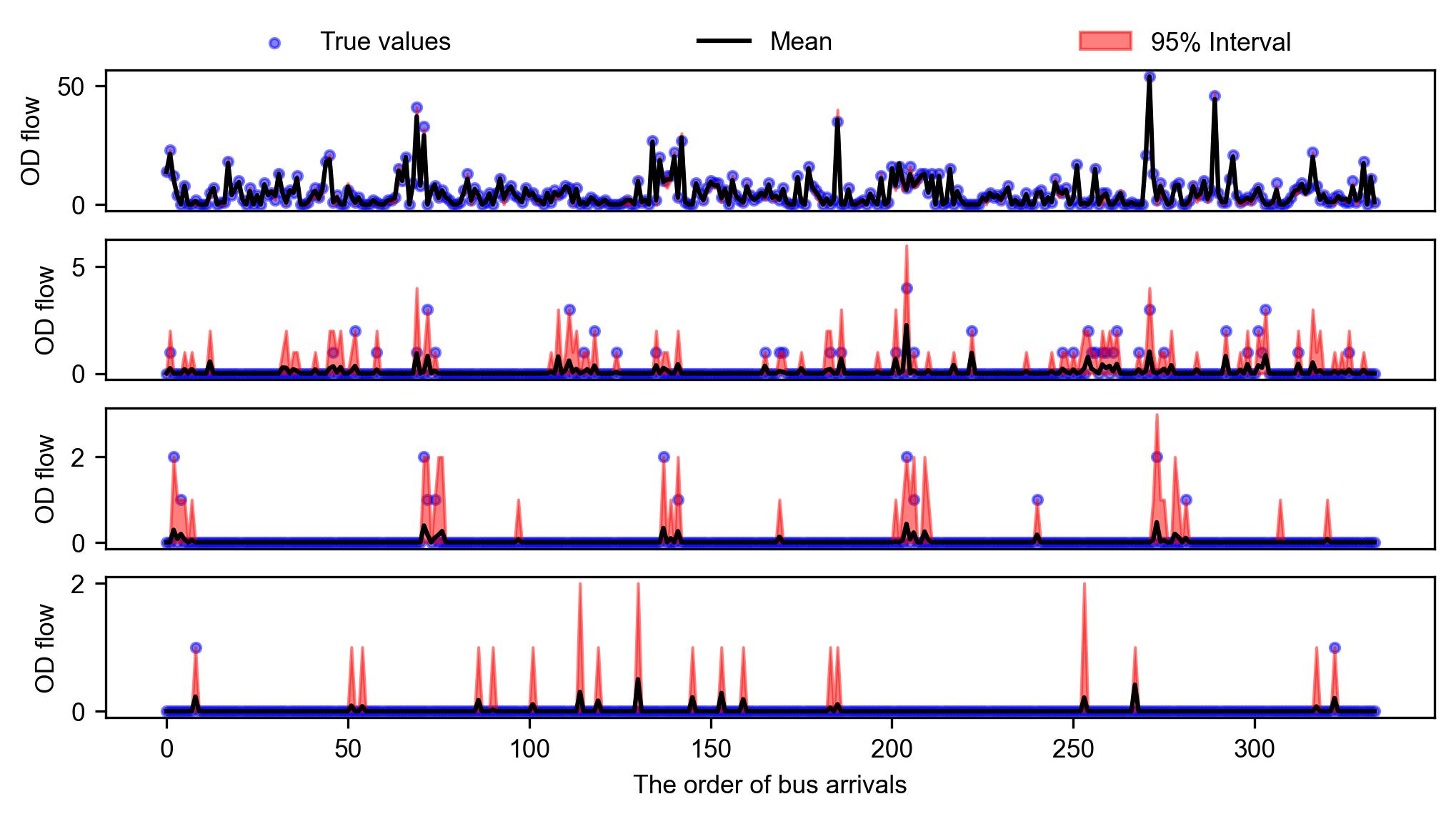}
\caption{Estimation with the uncertainty of some OD pairs on the long bus route.}
\label{fig:est_uncer_long}
\end{figure}

\end{document}